\documentclass[aps,prx,reprint,superscriptaddress]{revtex4-2}

\pdfoutput=1

\usepackage[utf8]{inputenc}
\usepackage{amsmath,amssymb,amsfonts,graphicx,verbatim,hyperref, mathtools, bbold, float}
\usepackage{physics}
\usepackage{soul, color}
\usepackage{algorithm, algorithmic}
\hypersetup{
	colorlinks=true,
	linkcolor=red,
	citecolor=blue,
	filecolor=blue,
}

\soulregister\ref{7}  % so that \hl and \st can wrap around \ref
\soulregister\cite{7} % so that \hl and \st can wrap around \cite
\newcommand{\J}{\mathcal{J}}
\newcommand*\diff{\mathop{}\!\mathrm{d}}

\newcommand*{\im}{\, \mathrm{Im}}
\newcommand*{\re}{\, \mathrm{Re}}
\newcommand*{\sgn}{\, \mathrm{sgn}}
\renewcommand*{\eqref}[1]{Eq.~(\ref{#1})}

\newcommand*{\figref}[1]{Fig.~\ref{#1}}

\newcommand*{\secref}[1]{Sec.~\ref{#1}}

\newcommand*{\appref}[1]{Appendix~\ref{#1}}
\newcommand*{\Appref}[1]{Appendix~\ref{#1}}
\newcommand*{\tabref}[1]{Tab.~\ref{#1}}

\begin{document}

\title{Energy Transport in Sachdev-Ye-Kitaev Networks Coupled to Thermal Baths}
\date{\today}
\author{Cristian \surname{Zanoci}}
\email{czanoci@mit.edu}
\affiliation{Department of Physics, Massachusetts Institute of Technology, 77 Massachusetts Avenue, Cambridge, Massachusetts 02139, USA}
\author{Brian \surname{Swingle}}
\email{bswingle@umd.edu}
\affiliation{Condensed Matter Theory Center and Joint Quantum Institute, Department of Physics, University of Maryland, College Park, Maryland 20742, USA}
\affiliation{Department of Physics, Brandeis University, Waltham, Massachusetts 02453, USA}

\begin{abstract}
We develop a general framework for studying the equilibrium and non-equilibrium properties of arbitrary networks of Sachdev-Ye-Kitaev clusters coupled to thermal baths. We proceed to apply this technique to the problem of energy transport, which is known to be diffusive due to the strange metal behavior of these models. We use the external baths to impose a temperature gradient in the system and study the emerging non-equilibrium steady state using the Schwinger-Keldysh formalism. We consider two different configurations for the baths, implementing either a boundary or bulk driving, and show that the latter leads to a significantly faster convergence to the steady state. This setup allows us to compute both the temperature and frequency dependence of the diffusion constant. At low temperatures, our results agree perfectly with the previously known values for diffusivity in the conformal limit. We also establish a relationship between energy transport and quantum chaos by showing that the diffusion constant is upper bounded by the chaos propagation rate at all temperatures. Moreover, we find a simple analytical form for the non-equilibrium Green's functions in the linear response regime and use it to derive exact closed-form expressions for the diffusion constant in various limits. We mostly focus on uniform one-dimensional chains, but we also discuss higher-dimensional generalizations.
\end{abstract}

\maketitle

\section{Introduction}
\label{sec:intro}

The subject of non-equilibrium dynamics spans across multiple branches of physics, ranging from condensed matter to quantum gravity. Recently, it has drawn a lot of attention in an attempt to answer fundamental questions regarding thermalization, quantum many-body chaos, and transport in strongly interacting systems. The latter is particularly useful for unveiling the properties of quantum matter and exploring the dynamical processes governing the behavior of quantum systems out of equilibrium. Despite recent efforts, practical calculations of transport coefficients in quantum many-body systems remain challenging from both a conceptual and technical standpoint~\cite{bertini2021}. On one hand, a general hydrodynamic description of transport is possible at a macroscopic level, but relating the parameters of this theory to the microscopic parameters of the underlying model is difficult except in special cases, for example, at weak coupling. On the other hand, numerical methods for simulating non-equilibrium dynamics can directly address a wide class of microscopic models, but their applicability is typically limited to small systems and short time scales. Therefore, it would be very useful to develop a framework that allows us to efficiently extract the transport properties of small systems, while simultaneously offering insights about the non-equilibrium dynamics in the hydrodynamic limit. 

In the context of one-dimensional models, open-system dynamics has shown promising results for bridging the gap between macroscopic effective theories and microscopic models, especially when combined with tensor network techniques~\cite{bertini2021,weimer2021,landi2021}. Within this framework, the system is coupled to external baths that drive the system towards a desired steady state, from which various transport properties are readily available. However, for a generic non-integrable model, these simulations can suffer from entanglement growth and very slow convergence at low temperatures~\cite{zanoci2021}. Moreover, very little is known about the structure of the emergent non-equilibrium states. Our goal is to study a solvable open-system model and show how one can mitigate the aforementioned convergence problems, while also developing an analytical understanding of the non-equilibrium steady state (NESS) and elucidating the relationship between transport and other many-body phenomena, such as quantum chaos. 

Looking more broadly, we expect the results about the structure of steady states and convergence rates in the solvable model to generalize to a variety of more physical models. Combined with tensor network methods, these lessons may significantly improve our ability to compute transport properties from microscopic theories. There are also a growing number of principled numerical techniques~\cite{Haegeman2011,leviatan2017quantum,white2018quantum,rakovszky2020dissipation,polkovnikov2010,schachenmayer2015}, besides the open-system approach, which might also benefit from additional information about the structure of current carrying states.

Our solvable model is built from the Sachdev-Ye-Kitaev (SYK) model~\cite{kitaev2015simple,sachdev1993,parcollet1999,sachdev2015,georges2000,georges2001,maldacena2016,kitaev2018,sarosi2017,rosenhaus2019}, which in recent years has emerged as a paradigmatic example of a strongly coupled, yet exactly solvable quantum many-body system. The model describes $N$ fermions with random all-to-all $q$-body interactions and displays a multitude of remarkable properties. The Hamiltonian belongs to a class of systems realizing holographic quantum matter without quasiparticle excitations~\cite{sachdev2015}. In the limit of large $N$ and low temperatures, the system has an emergent approximate time reparameterization conformal symmetry~\cite{kitaev2015simple,maldacena2016}. The SYK model also exhibits many-body chaos~\cite{kitaev2015simple,maldacena2016} and saturates a universal bound on the relevant Lyapunov exponent~\cite{maldacena2016chaos}, a feature shared by black holes in Einstein gravity. In fact, the SYK model is holographically dual to gravitational theories of black holes with near-horizon AdS$_2$ geometry~\cite{jakiw1985,teitelboim1983,sachdev2010holographic,sachdev2010strange,kitaev2015simple,almheiri2015,maldacena2016ads,kitaev2018,engelsy2016}. On the gravity side, this connection can be used to study unsolved questions related to holography and black holes, such as black hole evaporation~\cite{hawking1974,page1993,rocha2010,engelsy2016}. On the condensed matter side, the SYK model presents a valuable platform for studying non-Fermi liquid behavior~\cite{sachdev1993,parcollet1999,sachdev2015} and thermalization~\cite{deutsch1991,srednicki1994,sonner2017,garcia2018,haque2019,eberlein2017,bhattacharya2019,almheiri2019,zhang2019,haldar2020,kuhlenkamp2020}, among other phenomena. 

Subsequent works~\cite{gu2017diffusion,davison2017,song2017,patel2018,chowdhury2018,cmjian2017} extended the SYK model to higher spatial dimensions by coupling individual SYK clusters to their neighbors with similar SYK interactions, thus building lattices that exhibit strange metal behavior. The generalized models remain exactly solvable via a saddle point expansion and retain many features of the original SYK, including local criticality and maximal chaos~\cite{gu2017diffusion,davison2017}. The spatial locality of the models allows for the investigation of transport and chaos in these systems. This revealed many indicative properties of a strange metal, such as diffusive propagation of energy~\cite{gu2017diffusion,davison2017,song2017} and resistivity that scales linearly with temperature~\cite{song2017}. Moreover, it was shown that the same time-reparametrization field is responsible for the propagation of both low-energy modes and quantum chaos in this system~\cite{gu2017diffusion}, thus leading to a connection between the diffusion constant and the butterfly velocity~\cite{hartnoll2014,blake2016_1,blake2016_2,blake2017,hartman2017,choi2021,gu2017}.

Alongside   these   theoretical   developments, several experimental proposals for the SYK model have been put forward~\cite{franz2018,rahmani2019}, including realizations in ultracold atoms~\cite{danshita2017}, Majorana modes at the interface of a topological insulator and superconductor~\cite{pikulin2017}, semiconductor wires coupled through a disordered quantum dot~\cite{chew2017}, and graphene flakes with irregular boundaries~\cite{chen2018}. These experiments open up the possibility of directly studying the transport properties of the model. 

In this paper, we study energy transport in SYK networks driven out of equilibrium by coupling to external baths. The system-reservoir interaction in SYK models has been previously discussed in the context of thermalization~\cite{zhang2019,almheiri2019,haldar2020}, and only recently in the context of transport~\cite{cheipesh2020}. Most of the previous approaches to transport used the explicit form of the fermion four-point function and relied on the low-temperature or large $q$ limits to compute the diffusion constant~\cite{gu2017diffusion,gu2017,cmjian2017}. One advantage of coupling the system to baths is that it allows us to study transport for arbitrary values of $q$ and inverse temperature $\beta$. We consider two configurations with baths attached either on the boundary or throughout the bulk, and show that the latter leads to faster convergence. We study the resulting non-equilibrium dynamics using the Schwinger-Keldysh formalism~\cite{kamenev2011field,stefanucci2013nonequilibrium} and derive the associated Kadanoff-Baym equations governing the approach to a steady state. Once we reach the NESS, we compute the current and energy gradient across the system to determine the diffusion coefficient. 

We rely on this setup to investigate both DC and AC transport. For the former, the baths are used to impose a constant temperature imbalance on the system, while for the latter, we oscillate the temperature of the baths to create a time-dependent drive. For the DC case, we find that the diffusivity increases with $\beta$ until it saturates at a constant value matching exactly the conformal answer~\cite{gu2017diffusion}. We also show that the diffusion constant is related to chaos via the inequality $D\leq v_B^2/\lambda_L$, which holds at all temperatures and becomes an equality in the conformal limit. This behavior is consistent with studies of related systems~\cite{gu2017,lucas2016,hartman2017,choi2021}. For the AC case, we compute the frequency dependence of the diffusivity and find that it decreases exponentially at high frequencies. 

Additionally, our results indicate that the Green's functions in NESS are only slightly perturbed from their equilibrium values, allowing us to re-write the equations of motion in terms of these new non-equilibrium corrections without referencing the baths. In the case of DC transport, we were able to find this correction explicitly in terms of the equilibrium Green's function and derived analytic formulas for the diffusion constant in the limit of large $q$, as well as for $q=2$. In the case of AC transport, we found that the NESS contribution has a more complicated form which can only be estimated numerically. Our results computed via this formalism agree well with the ones obtained from coupling the system to baths and represent one of the first attempts at characterizing the NESS analytically. 

The rest of the paper is structured as follows. In \secref{sec:setup} we define our higher-dimensional generalization of the SYK model coupled to baths. We describe in detail our approach to non-equilibrium dynamics and show how to compute the different quantities related to transport.  In \secref{sec:equilibrium} we briefly review the equilibrium properties of our system. Next, in \secref{sec:dc} we present our results for DC transport on different lattices, while focusing specifically on the temperature dependence of the diffusion coefficients and the structure of the emerging NESS. Similarly, \secref{sec:ac} is devoted to frequency-dependent transport. Finally, we give a short discussion and outlook in \secref{sec:discussion}.

\section{Setup}
\label{sec:setup}
%General Framework / Formalism

We begin by introducing our family of models, the equations governing their equilibrium and non-equilibrium dynamics, the different approximations which significantly simplify these equations, and the general setup and relevant quantities used to study the transport properties of these models. 

\subsection{SYK models coupled to thermal baths}
\label{sec:models}

The SYK model~\cite{kitaev2015simple,sachdev1993,parcollet1999,georges2000,georges2001,sachdev2015,maldacena2016,kitaev2018,sarosi2017,rosenhaus2019} is a strongly interacting fermion model in $(0+1)$ dimensions. We consider a generalization of this model on arbitrary graphs $\mathcal{G}=(V, E)$. Each vertex $u\in V$ is an SYK cluster (quantum dot) of $N$ Majorana fermions $\psi_i^u$ with random all-to-all $q_S-$body interactions given by the Hamiltonian 
\begin{equation}
    H_0^u = i^{q_S/2}\sum_{\{i\}}J_{i_1\ldots i_{q_S}}^{(0)} \psi_{i_1}^u\ldots \psi_{i_{q_S}}^u, 
\end{equation}
where $\{i\}$ denotes the restricted sum over $1\leq i_1 < \cdots < i_{q_S}\leq N$. The vertices of the graph connected by an edge $(u, v)\in E$ are coupled to each other via the interaction Hamiltonian 
\begin{equation}
    H_1^{uv} = i^{q_I/2}\sum_{\{i\},\{j\}}J_{i_1\ldots i_{\frac{q_I}{2}}j_1\ldots j_{\frac{q_I}{2}}}^{(1)} \psi_{i_1}^u\ldots \psi_{i_{\frac{q_I}{2}}}^u\psi_{j_1}^v\ldots \psi_{j_{\frac{q_I}{2}}}^v. 
\end{equation}
The Majorana fermions obey the standard anti-commutation relations $\{\psi_i^u, \psi_j^v\}=\delta_{ij}\delta_{uv}$. Note that it is not essential that the interaction term contains the same number of fermions from both sites and a generalization to more generic interactions should be straightforward. Similar higher-dimensional SYK models have been previously studied in the context of transport~\cite{gu2017diffusion, gu2017,davison2017,song2017,patel2018,chowdhury2018,guo2019,can2019}, quantum chaos~\cite{gu2017entanglement,chen2017,bentsen2019}, and quantum phase transitions~\cite{banerjee2017,haldar2018,cmjian2017,jian2017mbl,cai2018}.

In order to study the non-equilibrium properties of our system, we couple a subset of vertices to thermal baths~\cite{chen2017,zhang2019,almheiri2019,cheipesh2020,haldar2020}. Each bath is modeled as an SYK cluster of $M$ Majorana fermions $\chi_i^u$ with all-to-all $q_B-$body interactions given by the Hamiltonian 
\begin{equation}
    H_B^u = i^{q_B/2}\sum_{\{i\}}J_{i_1\ldots i_{q_B}}^{(B)} \chi_{i_1}^u\ldots \chi_{i_{q_B}}^u. 
\end{equation}
The system-bath interaction is of the same form as the inter-cluster coupling on the graph
\begin{equation}
    H_{SB}^{u} = i^{q_I/2}\sum_{\{i\},\{j\}}V_{ i_1\ldots i_{\frac{q_I}{2}}j_1\ldots j_{\frac{q_I}{2}}}^{(u)} \psi_{i_1}^u\ldots \psi_{i_{\frac{q_I}{2}}}^u\chi_{j_1}^u\ldots \chi_{j_{\frac{q_I}{2}}}^u.
\end{equation}
Since not all vertices are necessarily coupled to a bath, we use $V^{(u)}\neq 0$ to indicate the presence of a bath and set $V^{(u)} = 0$ otherwise. The baths are coupled to the system at time $t=0$. 

All the SYK couplings are independent Gaussian random variables with zero mean and variances 

\begin{align}
    \langle (J_{i_1\ldots i_{q_S}}^{(0)})^2 \rangle &= \frac{J_0^2 (q_S-1)!}{N^{q_S-1}}, \\
    \langle (J_{i_1\ldots i_{q_I/2}j_1\ldots j_{q_I/2}}^{(1)})^2 \rangle &= \frac{J_1^2 (q_I/2)!(q_I/2-1)!}{N^{q_I-1}}, \\
    \langle (J_{i_1\ldots i_{q_B}}^{(B)})^2 \rangle &= \frac{J_B^2 (q_B-1)!}{M^{q_B-1}}, \\
    \langle (V_{ i_1\ldots i_{q_I/2}j_1\ldots j_{q_I/2}}^{(u)})^2 \rangle &= \frac{V_u^2 (q_I/2)!(q_I/2-1)!}{M^{q_I/2}N^{q_I/2-1}}.
\end{align}
The numerical coefficients are chosen to cancel additional factors in the path integral and the powers of $N$ ensure the correct scaling of extensive thermodynamic variables, such as the energy.

The total Hamiltonian is simply the sum of all the terms
\begin{equation}
    H = \sum_{u\in V} \big(H_0^u + H_B^u + H_{SB}^{u}\big) + \sum_{(u, v)\in E} H_1^{uv}.
    \label{eq:ham}
\end{equation}
Our model represents a unified framework for studying both equilibrium and non-equilibrium properties of arbitrary SYK lattices coupled to baths. We can reproduce many of the previously studied SYK configurations~\cite{gu2017diffusion,davison2017,song2017,patel2018,chowdhury2018,guo2019,can2019,gu2017entanglement,bentsen2019,banerjee2017,haldar2018,cmjian2017,jian2017mbl,cai2018,chen2017,zhang2019,almheiri2019,cheipesh2020,haldar2020,kuhlenkamp2020} by adjusting the topology of the graph, the type of intra- and inter-cluster interactions, or the nature of the baths in our model.  

\subsection{Equilibrium}
\label{sec:equilibrium_setup}

The generalized SYK model maintains all the exactly solvable properties of the original model in the large-$N$ limit~\cite{maldacena2016}. Strictly speaking, in the presence of quenched disorder, we would have to introduce replicas in the path integral. However, in the large-$N$ limit, our model self-averages and the interaction between replicas is suppressed~\cite{maldacena2016,gu2017diffusion}. Therefore, it is sufficient to consider the replica-diagonal partition function~\cite{maldacena2016,gu2017diffusion}, for which the Euclidean effective action can be written as
\begin{gather}
    S = \sum_{(u, v)\in E} S_{uv} + \sum_{u\in V} S_{u}, \\
    S_{uv} = - \frac{NJ_1^2}{q_I}\int \diff\tau_1 \diff\tau_2 G_u^S(\tau_1, \tau_2)^{q_I/2}
    G_v^S(\tau_1, \tau_2)^{q_I/2},
\end{gather}
\begin{widetext}
\begin{equation}
\begin{split}
    S_{u} &= -\frac{N}{2}\log\det \left(\partial_\tau - \Sigma_u^{S}(\tau_1, \tau_2)\right) -\frac{M}{2}\log\det \left(\partial_\tau - \Sigma_u^{B}(\tau_1, \tau_2)\right) - \frac{NV_u^2}{q_I}\int \diff\tau_1 \diff\tau_2 G_u^S(\tau_1, \tau_2)^{q_I/2}
    G_u^B(\tau_1, \tau_2)^{q_I/2} \\
    &\hspace{-0.4cm} + \frac{N}{2}\int \diff\tau_1 \diff\tau_2\left(\Sigma_u^S(\tau_1, \tau_2)G_u^S(\tau_1, \tau_2) - \frac{J_0^2}{q_S}G_u^S(\tau_1, \tau_2)^{q_S} \right) + \frac{M}{2}\int \diff\tau_1 \diff\tau_2\left(\Sigma_u^B(\tau_1, \tau_2)G_u^B(\tau_1, \tau_2) - \frac{J_B^2}{q_B}G_u^B(\tau_1, \tau_2)^{q_B} \right).
\end{split}
\end{equation}
\end{widetext}
For each vertex $u$, we introduced the Euclidean time-ordered fermion two-point functions 
\begin{align}
    G_u^S(\tau_1, \tau_2) &= \frac{1}{N} \sum_{i=1}^N \langle T\psi_i(\tau_1)\psi_i(\tau_2) \rangle,\\
    G_u^B(\tau_1, \tau_2) &= \frac{1}{M} \sum_{i=1}^M \langle T\chi_i(\tau_1)\chi_i(\tau_2) \rangle,
\end{align}
and the fermion self-energies $\Sigma_u^{S,B}$ as the associated Lagrange multipliers. In the large-$N$ limit, the saddle point of this effective action produces the Schwinger-Dyson (SD) equations of motion 
\begin{align}
    1 &= (\partial_\tau - \Sigma_u^{S, B}(\tau))G_u^{S,B}(\tau),\\
    \Sigma_u^S(\tau) &=  J_0^2G_u^S(\tau)^{q_S-1} + V_u^2 G_u^S(\tau)^{q_I/2-1}G_u^B(\tau)^{q_I/2}\nonumber \\
    &+ J_1^2\sum_{v\in V} A_{uv}G_u^S(\tau)^{q_I/2-1}G_v^S(\tau)^{q_I/2}, \label{eq:system_SD} \\
    \Sigma_u^B(\tau) &= J_B^2G_u^B(\tau)^{q_B-1} + \frac{N}{M}V_u^2G_u^S(\tau)^{q_I/2}G_u^B(\tau)^{q_I/2-1},
    \label{eq:bath_SD}
\end{align}
where $A_{uv}$ is the adjacency matrix of the graph and we assumed time-translation symmetry $\tau=\tau_1-\tau_2$ in equilibrium.

If the bath is much larger than the system $M\gg N$, we can neglect the back-reaction on the bath and drop the second term in \eqref{eq:bath_SD}. Hence, even though the smaller system is affected by its coupling to the bath through the term proportional to $V_u^2$, the larger bath can be approximated as decoupled~\cite{almheiri2019,zhang2019}. As a result, the bath's Green's function always matches its equilibrium value, even when coupled to the system. We will operate under this assumption in the remainder of the paper. 

\subsection{Non-equilibrium}
\label{sec:nonequilibrium_setup}

The real-time evolution of a quantum many-body system coupled to baths can be described using the Schwinger-Keldysh formalism~\cite{kamenev2011field,stefanucci2013nonequilibrium}. Following the derivation in Refs.~\cite{eberlein2017,bhattacharya2019}, we can write down the terms in the Lorentzian action

\begin{equation}
    S_{uv}  = \frac{iNJ_1^2}{q_I}\int_{\mathcal{C}} \diff t_1 \diff t_2 G_u^S(t_2, t_1)^{q_I/2}
    G_v^S(t_1, t_2)^{q_I/2},
\end{equation}
\begin{widetext}
\begin{align}
    S_{u} & = -\frac{iN}{2}\log\det \left(\partial_t\delta_{\mathcal{C}}(t_1, t_2) + i\Sigma_u^{S}(t_1, t_2)\right)+ \frac{iN}{2}\int_{\mathcal{C}} \diff t_1 \diff t_2\left( \frac{J_0^2}{q_S}G_u^S(t_2, t_1)^{q_S/2}G_u^S(t_1, t_2)^{q_S/2}-\Sigma_u^S(t_1, t_2)G_u^S(t_2, t_1) \right)\nonumber\\ &-\frac{iM}{2}\log\det \left(\partial_t\delta_{\mathcal{C}}(t_1, t_2) + i\Sigma_u^{B}(t_1, t_2)\right)+ \frac{iM}{2}\int_{\mathcal{C}} \diff t_1 \diff t_2\left( \frac{J_B^2}{q_B}G_u^B(t_2, t_1)^{q_B/2}G_u^B(t_1, t_2)^{q_B/2}-\Sigma_u^B(t_1, t_2)G_u^B(t_2, t_1) \right)\nonumber\\
    &+ \frac{iNV_u^2}{q_I}\int_{\mathcal{C}} \diff t_1 \diff t_2 G_u^S(t_2, t_1)^{q_I/2}
    G_u^B(t_1, t_2)^{q_I/2},
\end{align}
\end{widetext}
where $\mathcal{C} = \mathcal{C}^+ \cup \mathcal{C}^-$ is the closed-time Keldysh contour with the positive (forward) branch $\mathcal{C}^+$ going from $-\infty$ to $+\infty$ and the negative (backward) branch $\mathcal{C}^-$ going from $+\infty$ to $-\infty$. Recall that in the Schwinger-Keldysh formalism, the contour-ordered Green's functions are actually $2\times2$ matrices

%= -\frac{i}{N} \sum_{i=1}^N \langle T_{\mathcal{C}}\psi_i(t_1)\psi_i(t_2) \rangle
\begin{equation}
    G_u^{S}(t_1, t_2) = \begin{pmatrix}
    G_u^S(t_1^+, t_2^+) & G_u^S(t_1^+, t_2^-) \\
    G_u^S(t_1^-, t_2^+) & G_u^S(t_1^-, t_2^-)
    \end{pmatrix},
\end{equation}
where $t_i^+$ lives on the positive branch and $t_i^-$ lives on the negative branch. We will be mainly interested in the greater and lesser Green's functions

\begin{align}
    G_u^>(t_1, t_2) \equiv G_u^S(t_1^-, t_2^+) &= -\frac{i}{N} \sum_{i=1}^N \langle \psi_i(t_1^-)\psi_i(t_2^+) \rangle,\\
    G_u^<(t_1, t_2) \equiv G_u^S(t_1^+, t_2^-) &= -\frac{i}{N} \sum_{i=1}^N \langle \psi_i(t_1^+)\psi_i(t_2^-) \rangle,
\end{align}
through which we can define the other two components

\begin{align}
    G_u^{S}(t_1^+, t_2^+) &= \theta(t_1 - t_2)G_u^>(t_1, t_2) + \theta(t_2 - t_1)G_u^<(t_1, t_2),\nonumber\\
    G_u^{S}(t_1^-, t_2^-) &= \theta(t_1 - t_2)G_u^<(t_1, t_2) + \theta(t_2 - t_1)G_u^>(t_1, t_2).
    \label{eq:contour_identity}
\end{align}
In practice, it is often more convenient to work with the retarded, advanced, and Keldysh Green's functions

\begin{align}
    G_u^R(t_1, t_2) &= \Theta(t_1-t_2)\big( G_u^>(t_1, t_2)-G_u^<(t_1, t_2)\big),\\
    G_u^A(t_1, t_2) &= \Theta(t_2-t_1)\big( G_u^<(t_1, t_2)-G_u^>(t_1, t_2)\big),\\
    G_u^K(t_1, t_2) &= G_u^>(t_1, t_2)+G_u^<(t_1, t_2),
\end{align}
which are related to the contour-ordered Green's functions above via a Keldysh rotation~\cite{kamenev2011field,stefanucci2013nonequilibrium}. The corresponding self-energies are defined in a similar manner. For Majorana fermions in thermal equilibrium, the greater and lesser Green's functions are related via
\begin{equation}
    G_u^>(t_1, t_2) =  -G_u^<(t_2, t_1) = -G_u^>(t_2, t_1)^*,
    \label{eq:gtr_lsr}
\end{equation}
which holds true even out-of-equilibrium as long as the time evolution starts from thermal equilibrium~\cite{babadi2015}. Therefore, all other two-point functions can be computed from $G_u^>(t_1, t_2)$ and we can focus solely on studying this component of the Green's function. Furthermore, due to the Majorana anti-commutation relations, the greater Green's function at equal time reduces to $G_u^>(t, t) =  -i/2$.

%To obtain the Schwinger-Dyson equations in Lorentzian time, we can either use the large-$N$ saddle point equations for the Lorentzian action~\cite{eberlein2017,bhattacharya2019} or perform an analytic continuation of the equilibrium SD equations from imaginary to real time~\cite{almheiri2019}, yielding 
To obtain the Schwinger-Dyson equations in real time, we can use the large-$N$ saddle point equations for the Lorentzian action~\cite{eberlein2017,bhattacharya2019}, yielding 
\begin{align}
    \Sigma_u^{S, B}(t_1, t_2) &= \partial_t\delta_{\mathcal{C}}(t_1, t_2)-(G_u^{S, B}(t_1, t_2))^{-1},\\
    \Sigma_u^>(t_1, t_2) &= -i^{q_S}J_0^2G_u^>(t_1, t_2)^{q_S-1} \nonumber \\
    &\hspace{-1cm} -i^{q_I}V_u^2 G_u^>(t_1, t_2)^{q_I/2-1}G_{u, B}^>(t_1, t_2)^{q_I/2} \label{eq:self_energy} \\
    &\hspace{-1cm} -i^{q_I}J_1^2\sum_{v\in V} A_{uv}G_u^>(t_1, t_2)^{q_I/2-1}G_v^>(t_1, t_2)^{q_I/2}, \nonumber\\
    \Sigma_{u, B}^>(t_1, t_2) &= -i^{q_B}J_B^2G_{u, B}^>(t_1, t_2)^{q_B-1}.
\end{align}
Furthermore, one can apply convolutions on either side of the first SD equation, use the Langreth rules and integration by parts~\cite{kamenev2011field,stefanucci2013nonequilibrium,eberlein2017,bhattacharya2019}, to obtain the Kadanoff-Baym (KB) equations for the system's Green's functions

\begin{align}
    i\partial_{t_1}G_u^>(t_1, t_2) &= \nonumber \\
    &\hspace{-2cm}\int_{-\infty}^{\infty}\diff t_3\left(\Sigma_u^R(t_1, t_3)G_u^>(t_3, t_2) + \Sigma_u^>(t_1, t_3)G_u^A(t_3, t_2) \right),\\
    -i\partial_{t_2}G_u^>(t_1, t_2) &= \nonumber \\  &\hspace{-2cm}\int_{-\infty}^{\infty}\diff t_3\left(G_u^R(t_1, t_3)\Sigma_u^>(t_3, t_2) + G_u^>(t_1, t_3)\Sigma_u^A(t_3, t_2) \right).
\end{align}
In order to make the causal structure of the KB equations more explicit and for future use in our numerical calculations, we use the definitions of advanced and retarded propagators to write them as 

\begin{widetext}
\begin{equation}
\begin{split}
    i\partial_{t_1}G_u^>(t_1, t_2) &= \int_{-\infty}^{t_1}\diff t_3\left(\Sigma_u^>(t_1, t_3) + \Sigma_u^>(t_3, t_1) \right)G_u^>(t_3, t_2) - \int_{-\infty}^{t_2}\diff t_3 \Sigma_u^>(t_1, t_3)\left(G_u^>(t_3, t_2) + G_u^>(t_2, t_3)\right), \\
    -i\partial_{t_2}G_u^>(t_1, t_2) &= \int_{-\infty}^{t_1}\diff t_3\left(G_u^>(t_1, t_3) + G_u^>(t_3, t_1) \right)\Sigma_u^>(t_3, t_2) - \int_{-\infty}^{t_2}\diff t_3 G_u^>(t_1, t_3)\left(\Sigma_u^>(t_3, t_2) + \Sigma_u^>(t_2, t_3)\right). 
    \label{eq:KB}
\end{split}
\end{equation}
\end{widetext}
Thus the evolution of $G_u^>(t_1, t_2)$ only depends on $G_u^>(t, t')$ evaluated at earlier times $t < t_1$ and $t'< t_2$, making the causal structure evident. 

These equations are applicable to any non-equilibrium situations. They describe the generic time evolution of a Green's function, with the only information about the physical system encoded through the definition of self-energy in \eqref{eq:self_energy}. In our case, we couple the system to baths at different temperatures and observe the steady states that arise at late times.

The KB equations are generally not tractable analytically and one typically has to resort to numerical integration. Our approach is similar to the one used in Refs.~\cite{eberlein2017,guo2019,bhattacharya2019,almheiri2019,zhang2019,kuhlenkamp2020,haldar2020}. We begin by solving the SD equations self-consistently in Lorentzian time for each decoupled cluster~\cite{eberlein2017,guo2019}. This gives us the initial equilibrium conditions for $G_u^>(t_1, t_2)$ at all times $t_1, t_2 < 0$. Then, at $t_1=t_2=0$, we turn on the bath couplings $V_u$ and solve the KB equations on a discrete grid with spacing $\diff t=0.05$ (in units of inverse coupling) in the $(t_1, t_2)$ plane. The grid size is usually set to $1000L\times1000L$, where $L$ is the size of the system. In the case of frequency-dependent driving, the time step is further decreased such that $\omega \diff t\ll 1$ and the sub-sampling effects are negligible. The equations are solved using a predictor-corrector integration scheme~\cite{butcher2016,haldar2020}. Since the Green's functions typically decay exponentially away from the diagonal $t_1=t_2$, the calculations can be sped up by restricting our attention to a strip $|t_1-t_2|\lesssim 10\beta$ and setting all the Green's functions to zero outside that strip~\cite{almheiri2019}. Furthermore, \eqref{eq:gtr_lsr} reduces the computation time in half because we only have to solve the KB equations for $G_u^>(t_1, t_2)$ with $t_1>t_2$. 

Although the KB equations do not have a closed-form solution, they simplify significantly in the conformal and large $q$ limits, which we discuss next. 

\subsection{Conformal limit}
\label{sec:conformal}

%$\tau J_0, tJ_0 \gg 1$
At low temperatures $\beta J_0 \gg 1$ and large timescales $\tau J_0 \gg 1$, the system develops an emergent conformal symmetry~\cite{maldacena2016}. In Euclidean time, the Green's function for an isolated SYK cluster can be found explicitly 

\begin{equation}
    G_u^S(\tau) = b\left(\frac{\pi}{\beta \sin(\pi\tau/\beta)}\right)^{2/q_S}\sgn(\tau),
\end{equation}
where $b$ is given by the equation

\begin{equation}
    b^{q_S} = \frac{1}{\pi J_0^2}\left(\frac{1}{2} - \frac{1}{q_S}\right)\tan(\pi/q_S).
    \label{eq:conformal_b}
\end{equation}
Many properties of the SYK system, such as the fermion four-point function and the chaos exponent, can be computed exactly in this limit~\cite{maldacena2016}. We will use the conformal answer when studying the equilibrium properties of our system in \secref{sec:equilibrium}. 
%By analytic continuation to real time, we can obtain the greater Green's function

%\begin{equation}
%   G_u^>(t) = -ib\left(\frac{\pi}{\beta \sinh(\pi t/\beta)}\right)^{2/q_S}e^{-i\pi/q_S}.
%\end{equation}

\subsection{Large \texorpdfstring{$q$}{q} limit}
\label{sec:large_q}

In order to obtain an analytic approximation for the Green's function at both small and large energies, we take the large $q$ limit~\cite{maldacena2016,tarnopolsky2019}. For the sake of simplicity, we set $q_S=q_I=q_B=q$ throughout this section. To leading order in $1/q$, the system's Green's function and self-energy are given by

%\begin{align}
%    G_u^S(\tau) &= \frac{1}{2}\left(1+\frac{g_u(\tau)}{q}+\cdots\right)\sgn(\tau)\approx \frac{1}{2} e^{\frac{g_u(\tau)}{q}}\sgn(\tau),\\
%    \Sigma_u^S(\tau) &= \frac{1}{q} \Big( \J_0^2 e^{g_u(\tau)}+\mathcal{V}_u^2 e^{\frac{1}{2}(g_u(\tau)+g_u^B(\tau))}  \nonumber \\
%    &+ \J_1^2 e^{g_u(\tau)/2}\sum_v A_{uv}e^{g_v(\tau)/2}\Big)\sgn(\tau),
%\end{align}
\begin{align}
    G_u^>(t_1, t_2) &= -\frac{i}{2}\left(1+\frac{g_u(t_1, t_2)}{q}+\cdots\right)\approx -\frac{i}{2} e^{\frac{g_u(t_1, t_2)}{q}}, \\
    \Sigma_u^>(t_1, t_2) &= -\frac{i}{q}\Big(\J_0^2 e^{g_u(t_1, t_2)}+\mathcal{V}_u^2 e^{\frac{1}{2}(g_u(t_1, t_2)+g_u^B(t_1, t_2))} \nonumber \\
    &+ \J_1^2 e^{g_u(t_1, t_2)/2}\sum_v A_{uv}e^{g_v(t_1, t_2)/2}\Big),
\end{align}
where ``\ldots" denotes higher order terms in $1/q$, $g_u(t_1, t_2)$ is a function of order one satisfying $g_u(t, t) = 0$, and $\J_{0,1} = J_{0,1}\sqrt{q2^{1-q}}$, $\mathcal{V}_u = V_u\sqrt{q2^{1-q}}$ are re-scaled couplings. The large $q$ limit is well defined only when we adjust the original couplings ($J_0, J_1, V_u$) such that the re-scaled couplings ($\J_0, \J_1, \mathcal{V}_u$) are kept fixed as $q\to\infty$. 

We can use these expressions to write the KB equations in terms of the new functions $g_u(t_1, t_2)$~\cite{eberlein2017}. To leading order in $1/q$, we have

\begin{align}
    \partial_{t_1}g_u(t_1, t_2) &= 2iq\int_{-\infty}^{t_2}\diff t_3 \Sigma_u^>(t_1, t_3) \nonumber \\
    &\hspace{-0.1cm}- iq\int_{-\infty}^{t_1}\diff t_3\left(\Sigma_u^>(t_1, t_3) + \Sigma_u^>(t_3, t_1)\right), \label{eq:KB_1_large_q} \\
    \partial_{t_2}g_u(t_1, t_2) &= 2iq\int_{-\infty}^{t_1}\diff t_3 \Sigma_u^>(t_3, t_2) \nonumber \\
    &\hspace{-0.1cm}- iq\int_{-\infty}^{t_2}\diff t_3\left(\Sigma_u^>(t_3, t_2) + \Sigma_u^>(t_2, t_3)\right).
    \label{eq:KB_2_large_q}
\end{align}
Notice that in the first equation, $t_2$ only appears in the integration limit on the right-hand side (and similarly for $t_1$ in the second equation). This is a feature of the large $q$ expansion and it does not hold more generally. Therefore, the second time derivative of either equation takes on a simple form 

\begin{equation}
    \frac{\partial^2 g_u(t_1, t_2)}{\partial t_1 \partial t_2} = 2iq\Sigma_u^>(t_1, t_2).
\end{equation}
If we additionally assume time translation invariance, then we have

\begin{equation}
\begin{split}
    -\frac{\partial^2 g_u(t)}{\partial t^2} &=  2\Big(\J_0^2 e^{g_u(t)}+\mathcal{V}_u^2 e^{\frac{1}{2}(g_u(t)+g_u^B(t))} \\
    &+ \J_1^2 e^{g_u(t)/2}\sum_v A_{uv}e^{g_v(t)/2}\Big).
    \label{eq:KB_large_q_trans_inv}
\end{split}
\end{equation}
This results in a system of coupled ODEs that can be solved numerically. For a single isolated cluster, the solution can be obtained from the imaginary time formula derived in Ref.~\cite{maldacena2016} by analytic continuation

%= \left(\frac{1}{\cosh(\frac{\pi vt}{\beta}) + i\tan(\frac{\pi v}{2})\sinh(\frac{\pi vt}{\beta})}\right)^2,
\begin{equation}
    e^{g^{(0)}(t)} = \frac{\cos^2\left(\pi v/2\right)}{\cosh^2\left(\frac{\pi v}{\beta} \left(\frac{i\beta}{2}+t\right)\right)},
    \label{eq:g_0}
\end{equation}
where $v$ satisfies

\begin{equation}
     \beta \J_0 = \frac{\pi v}{\cos(\pi v/2)}.
     \label{eq:v}
\end{equation}

As shown in Ref.~\cite{eberlein2017}, given the symmetries of the KB equation and equilibrium initial conditions (see \eqref{eq:gtr_lsr}), the large $q$ solution always obeys

\begin{equation}
    g_u(-t) = g_u(t)^*,
    \label{eq:large_q_sym}
\end{equation}
which is indeed true for the equilibrium solution $g^{(0)}(t)$.

\subsection{Observables}
\label{sec:obs}

\begin{figure*}[ht]
	\begin{center}
	\includegraphics[width = \textwidth]{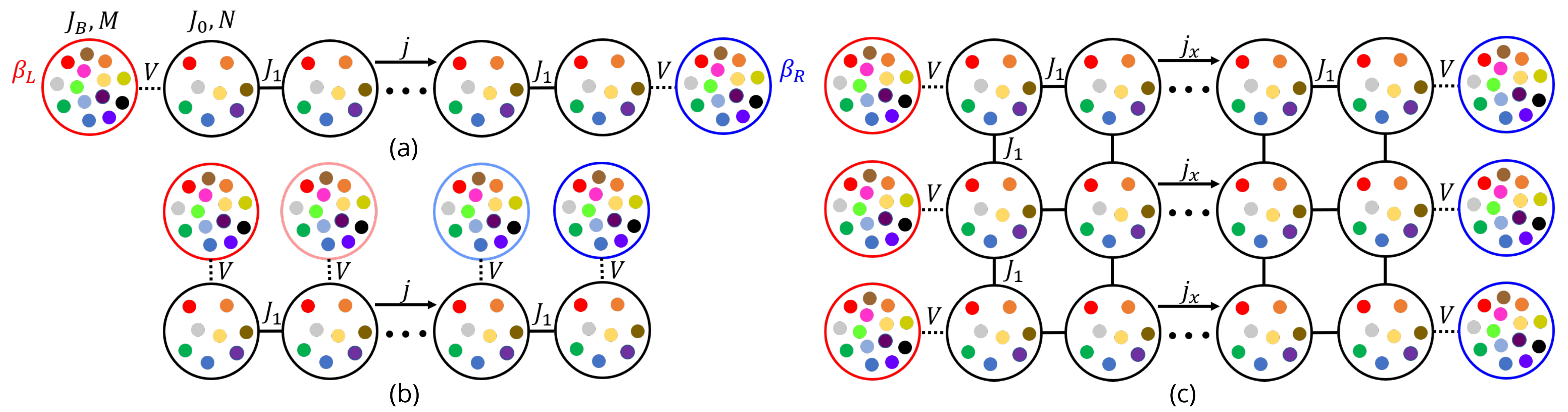}
	\caption{Schematic diagram of the non-equilibrium setups under study. (a) The boundary-driven SYK chain is connected at both ends to thermal baths at inverse temperatures $\beta_L$ and $\beta_R$. Each system site contains $N$ Majorana fermions and the hopping between neighboring sites is set by $J_1$, while the baths consist of $M$ Majorana fermions coupled to the chain with interaction $V$. In NESS, a homogeneous current $j$ flows through the bulk. (b) Same as (a), but with baths attached at every site throughout the chain. The temperature of the baths varies linearly from left to right. (c) Two-dimensional SYK lattice connected to baths at its two vertical edges, resulting in a net uniform horizontal current $j_x$.}
	\label{fig:fig1}
	\end{center}
\end{figure*}

Once we have access to the greater Green's functions, we can easily compute any observable in our system. We are interested in the transport of conserved quantities and in the case of the SYK Hamiltonian only the energy is conserved. It is convenient to decompose the system Hamiltonian in terms of bond operators $H_S=\sum_{(u, v)\in E} H^{uv} = \sum_{(u, v)\in E}(H_0^u/d_u+H_0^v/d_v+H_1^{uv})$ acting on sites $(u, v)$, where $d_u = \sum_{u'} A_{uu'}$ is the degree of vertex $u$. We then define separately the on-site energy per particle of each cluster

\begin{equation}
    E_0^u =  -i^{q_S+1}\frac{J_0^2}{q_S}\int_{-\infty}^{t}\diff t_1 \left(G_u^>(t, t_1)^{q_S} - G_u^>(t_1, t)^{q_S}\right),
\end{equation}
and similarly the interaction energy between clusters

\begin{align}
    E_1^{uv} &= -i^{q_I+1}\frac{2J_1^2}{q_I}\int_{-\infty}^{t}\diff t_1 \Big(G_u^>(t, t_1)^{q_I/2}G_v^>(t, t_1)^{q_I/2} \nonumber \\
    &- G_u^>(t_1, t)^{q_I/2}G_v^>(t_1, t)^{q_I/2}\Big).
\end{align}
The on-bond energy then becomes $E_{uv}\equiv\langle H^{uv}\rangle=E_0^u/d_u+E_0^v/d_v+E_1^{uv}$. The formula for the associated local energy current $j_u$ per particle flowing from $u$ to $v$ can be derived by combining the continuity equation at site $u$ with Heisenberg's equation of motion~\cite{zotos1997transport}

\begin{equation}
    j_u = i\sum_{u'}A_{u'u}[H^{u'u}, H^{uv}].
\end{equation}
Computing the expectation value of these commutators in the Schwinger-Keldysh formalism is more complicated. We provide a detailed derivation in \appref{sec:appendixA} and our general formula for the current is given by \eqref{eq:current_general}. 

In this paper, we will mainly be interested in regular lattices (one- and two-dimensional) with baths acting on the boundaries (see \figref{fig:fig1}). The right and left baths are held at different inverse temperatures $\beta_{R, L} = \beta \pm \delta\beta$, where $\beta$ is the average bath temperature and $\delta\beta$ is a  small temperature imbalance driving the system out of equilibrium. We typically choose $\delta\beta = 0.1\beta$ to make sure we are in the linear response regime. In the long-time limit, when the system reaches its NESS, the energy current becomes uniform in the bulk $j\equiv\langle j_u\rangle$. For the particular case of a uniform one-dimensional chain (\figref{fig:fig1}(a)), the current reads

\begin{equation}
    j = \frac{i^{q_S+q_I}}{2}J_1^2J^2\Re(j_{++}^{x-1,x,x+1}+j_{+-}^{x-1,x,x+1}),
    \label{eq:current_1D}
\end{equation}
where we labeled the sites by their horizontal coordinate $x$. Similarly, the energy gradient is simply the difference in energy on two consecutive bonds 

\begin{equation}
    \nabla E = E_{x, x+1} - E_{x-1, x}.
    \label{eq:gradient_1D}
\end{equation}

For a uniform setup with $q_S=q_I$, our model is a diffusive metal~\cite{gu2017diffusion}. We expect that transport in such a system is governed by Fourier's law $j=-D\nabla E=-D\Delta E/L$ with a temperature-dependent diffusion constant $D$. Here $\Delta E$ denotes the energy difference across the system and $L$ is the linear size of the system. More generally, a system can exhibit non-diffusive transport where the current scales with system size as $j\sim 1/L^\gamma$ with $\gamma\neq1$~\cite{bertini2021}. We explicitly verify in our numerical simulations that the transport is indeed diffusive at all temperatures, except for the case $q_S<q_I$, when the system becomes an insulator at low temperatures (see \secref{sec:other_models}). 

In order to accurately describe the temperature dependence of the diffusion constants, we need to introduce an effective local temperature. This temperature can in principle be very different from the average bath temperature at which we drive the system~\cite{zanoci2021,zanoci2016}. At sufficiently long times after turning on the bath couplings, the system reaches a nearly-thermal state where the Green's functions become time-translation invariant. This allows us to use the fluctuation-dissipation theorem (FDT)~\cite{kamenev2011field,stefanucci2013nonequilibrium,eberlein2017} to define an effective temperature $\beta_u^{-1}$ similarly to the thermal equilibrium case  

\begin{equation}
    \frac{iG_u^K(\omega)}{A_u(\omega)} = \tanh(\frac{\beta_u\omega}{2}),
    \label{eq:fdt}
\end{equation}
where $A_u(\omega) = -2\Im G_u^R(\omega)$ is the spectral function and $G_u^K(\omega)$ is the Fourier transform of the Keldysh Green's function. Note that the FDT holds only for low frequencies, since high frequencies are affected by the size of the discretization step $\diff t$. One can extract the local temperature by fitting \eqref{eq:fdt} over a small frequency interval~\cite{eberlein2017,bhattacharya2019}. Alternatively, we can access the low-frequency limit directly by taking the slope of the ratio of Green's functions at $\omega = 0$~\cite{zhang2019,cheipesh2020}

\begin{equation}
    \beta_u =\frac{\diff}{\diff\omega}\left( \frac{2iG_u^K(\omega)}{A_u(\omega)}\right)\bigg\rvert_{\omega=0}.
    \label{eq:local_temp}
\end{equation}
This simple relation unambiguously defines the local temperature of an SYK cluster, in contrast with conventional spin systems, where defining a non-equilibrium temperature is more challenging~\cite{zanoci2021}. Since the system is many-body chaotic, we expect it to reach a nearly-thermal state relatively fast, with a local temperature closely matching the bath temperature $\beta$~\cite{almheiri2019,zhang2019}. 

Finally, we would like to study how fast we approach the non-equilibrium steady state. This would allow us to compare the effectiveness of different driving setups. The convergence time to NESS can be defined for any of our observables, such as the current or the diffusion constant. Suppose we measure an observable $z_t$ at each step $t$ during our time evolution after coupling the system to the baths. To assess the convergence of a sequence of measurements $\{z_t\}_{t=0}^T$, we define the convergence time in terms of the error relative to the asymptotic value $z_\infty=\lim\limits_{T\to\infty} z_T$ of a given observable 

\begin{equation}
    \epsilon(t) = \left|1-\frac{z_t}{z_\infty} \right|.
    \label{eq:error}
\end{equation}
Note that this metric is invariant under rescaling, but not under a translation by a constant of the data. At late times, we expect the convergence to be exponentially fast $\epsilon(t)\sim e^{-\Gamma t}$, where $\Gamma$ is the convergence rate~\cite{almheiri2019}. We can also define a convergence time $t_{\mathrm{NESS}}$ as the minimum time such that $\epsilon(t)\leq\eta$ for all $t\geq t_{\mathrm{NESS}}$, where $\eta$ is a set threshold. Notice that the definition of $t_{\mathrm{NESS}}$ does not assume exponential convergence to NESS and is therefore more general. By comparing convergence times, we can quantitatively assess whether a certain bath configuration is more efficient at inducing the non-equilibrium steady state. In \secref{sec:bulk_dc} we will show that this is indeed the case for bulk driving.

\section{Equilibrium}
\label{sec:equilibrium}

We begin by studying the equilibrium setup, where all the bath and system clusters are at the same inverse temperature $\beta$. This will help us understand the changes that a local Green's function undergoes by simply connecting to its neighbors and the baths, before even introducing non-equilibrium effects caused by temperature imbalance. Since there is no dynamics, we look for solutions of the SD equations in Euclidean time. We will focus on the case $q_S=q_I=q_B=q$, but the more general setup was thoroughly discussed in Refs.~\cite{almheiri2019,zhang2019}. This case corresponds to a marginal interaction, such that the effective $q$ does not change, while the effective on-site interaction $J_0$ gets renormalized to $J_u$. In other words, the SD equation for an interacting SYK cluster has the exact same form as that of an isolated $(0+1)$ dimensional SYK model with coupling $J_u$

\begin{equation}
    \Sigma_u^S(\tau) = J_u^2G_u^S(\tau)^{q-1}.
\end{equation}

We will consider the conformal limit for simplicity, but the same qualitative results can be obtained by solving the SD equations numerically using the approach described in Ref.~\cite{maldacena2016} with minor adjustments to the self-energy formula as given by \eqref{eq:system_SD}. If $q$ is the same throughout the system, then the time-dependence cancels completely in the SD equations in the conformal limit. From \eqref{eq:conformal_b}, it is easy to see that the Green's function depends on the coupling as $G_u^S\sim J_u^{-2/q}$ and $\Sigma_u^S\sim J_u^{2/q}$. Therefore, \eqref{eq:system_SD} simplifies to 
    
\begin{equation}
    \frac{J_0^2}{J_u^2} + \frac{J_1^2}{J_u} \sum_{v}\frac{A_{uv}}{J_v} + \frac{V_u^2}{J_BJ_u}= 1,
    \label{eq:SD_equilibrium}
\end{equation}
for each site $u$. This is a system of quadratic equations for the effective coupling $J_u$, whose general solution can always be found numerically. 

\begin{figure}[t]
	\begin{center}
	\includegraphics[width = \columnwidth]{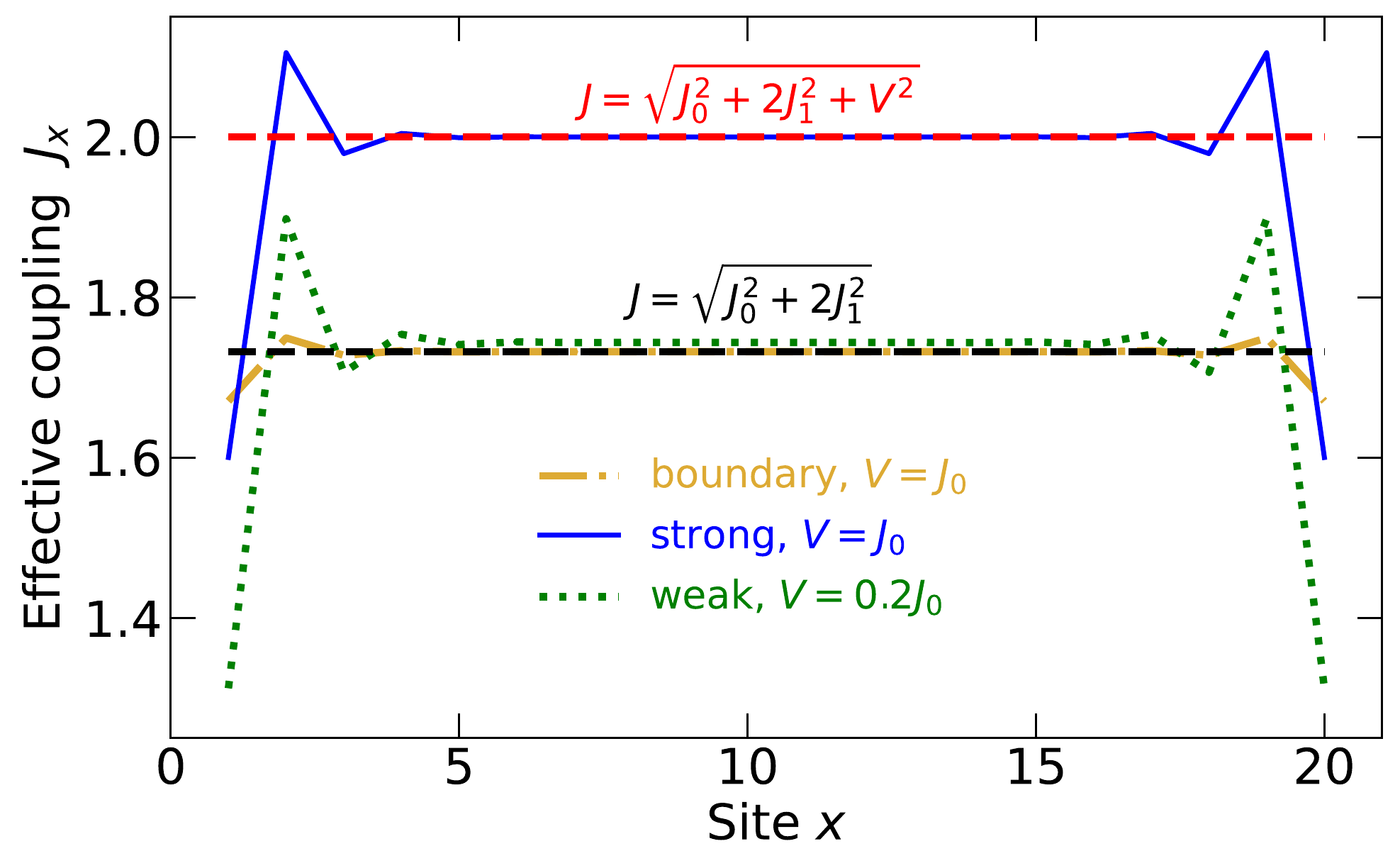}
	\caption{Equilibrium properties of an SYK chain of length $L=20$. The effective on-site coupling $J_x$ is shown as a function of system site in the case of boundary driving (dash-dotted line), strong bulk driving (solid line), and weak bulk driving (dotted line). The asymptotic values for the coupling deep in the bulk are shown with dashed lines. We set $J_0=J_1=1$.}
	\label{fig:fig2}
	\end{center}
\end{figure}

We further specialize to the case of a one-dimensional chain of $L$ SYK clusters, labeled by their lattice position $x$, with nearest-neighbor couplings and open boundary conditions (i.e. the only non-zero entries of the adjacency matrix are $A_{x, x-1} = A_{x, x+1} = 1$). We then investigate configurations in which the baths are either coupled at the two ends of the chain (boundary driving, see \figref{fig:fig1}(a)) or at every site of the chain (bulk driving, \figref{fig:fig1}(b)). The system-bath coupling always has the same value $V$. This setup is appropriate for studying the transport properties of a one-dimensional SYK system~\cite{gu2017diffusion, gu2017,davison2017,song2017,patel2018,chowdhury2018,guo2019,can2019} and can be easily generalized to higher dimensions (\figref{fig:fig1}(c)). 

Our results for the effective coupling $J_x$ obtained by solving the system of equations~(\ref{eq:SD_equilibrium}) is shown in \figref{fig:fig2}. In the case of boundary driving, we see a small edge effect due to the presence of the baths, but otherwise the effective coupling approaches its site-independent value $J=\sqrt{J_0^2+2J_1^2}$ deep in the bulk. Note that this effective coupling is slightly different from the one derived in Ref.~\cite{gu2017diffusion} due to a distinct choice of normalization for the variance of the coupling constant $J_1$. When baths are present on every site, their interaction with the system will strongly renormalize the coupling throughout the chain. For example, if we choose $J_B=J$, then the effective coupling away from the boundaries converges to $J=\sqrt{J_0^2+2J_1^2+V^2}$, which is very different from the boundary-driven case. In non-equilibrium, this bulk driving would significantly alter the transport coefficients.

We consider two approaches for mitigating the unwanted effects of bulk driving, while preserving some of its appealing features, such as fast NESS convergence rates. First, we could adjust the couplings $\tilde{V}=\tilde{J}_0=J_0/\sqrt{2}$ such that the effective on-site interaction in the bulk becomes the same as in the case of boundary driving $J=\sqrt{J_0^2+2J_1^2}$. We refer to this as strong bulk driving. The issue with this solution is that it slightly changes the properties of the underlying system by reducing $J_0$. Alternatively, we could take the limit of weak bulk driving $V\ll J_0, J_1$, which leaves the effective coupling mostly unchanged (see \figref{fig:fig2}). However, the rate at which energy is exchanged with the baths will also decrease, leading to slightly longer convergence times. The weak driving scheme serves as a middle ground between the strong and boundary drivings. In the next section, we will further explore the advantages of each setup in the context of non-equilibrium dynamics.

\section{DC Transport}
\label{sec:dc}

We now proceed with our results for the non-equilibrium case in the presence of time-independent (DC) driving. Following the procedure described in \secref{sec:setup}, we couple the system to baths at different temperatures and measure the energy gradients, local temperatures, and currents that arise as a result. We first consider the case of boundary driving and investigate the properties of the emerging NESS. We then show that bulk driving leads to approximately the same diffusion constants, but with much faster convergence times. Based on these findings, we postulate a simple ansatz for the non-equilibrium Green's function and show that it perfectly captures the transport properties of the SYK system. Using this ansatz, we derive a closed-form solution for the temperature-dependent diffusion coefficient in the large $q$ limit.  Throughout these sections, we focus on one-dimensional chains and set $q_S=q_I=q_B=q$ for simplicity. We discuss the case with $q_S\neq q_I$ separately in \secref{sec:other_models} and conclude our analysis of DC transport with a generalization to higher-dimensional lattices. To emphasize that our methods are applicable to a wide range of parameters, we display results for various values of $\beta$ and $q$. We also fix $\J_0=\J_1=1$ in order to treat the results for different $q$ on equal footing.

\subsection{Boundary-driven SYK chain}
\label{sec:boundary_dc}

\begin{figure}[tp]
	\begin{center}
	\includegraphics[width = \columnwidth]{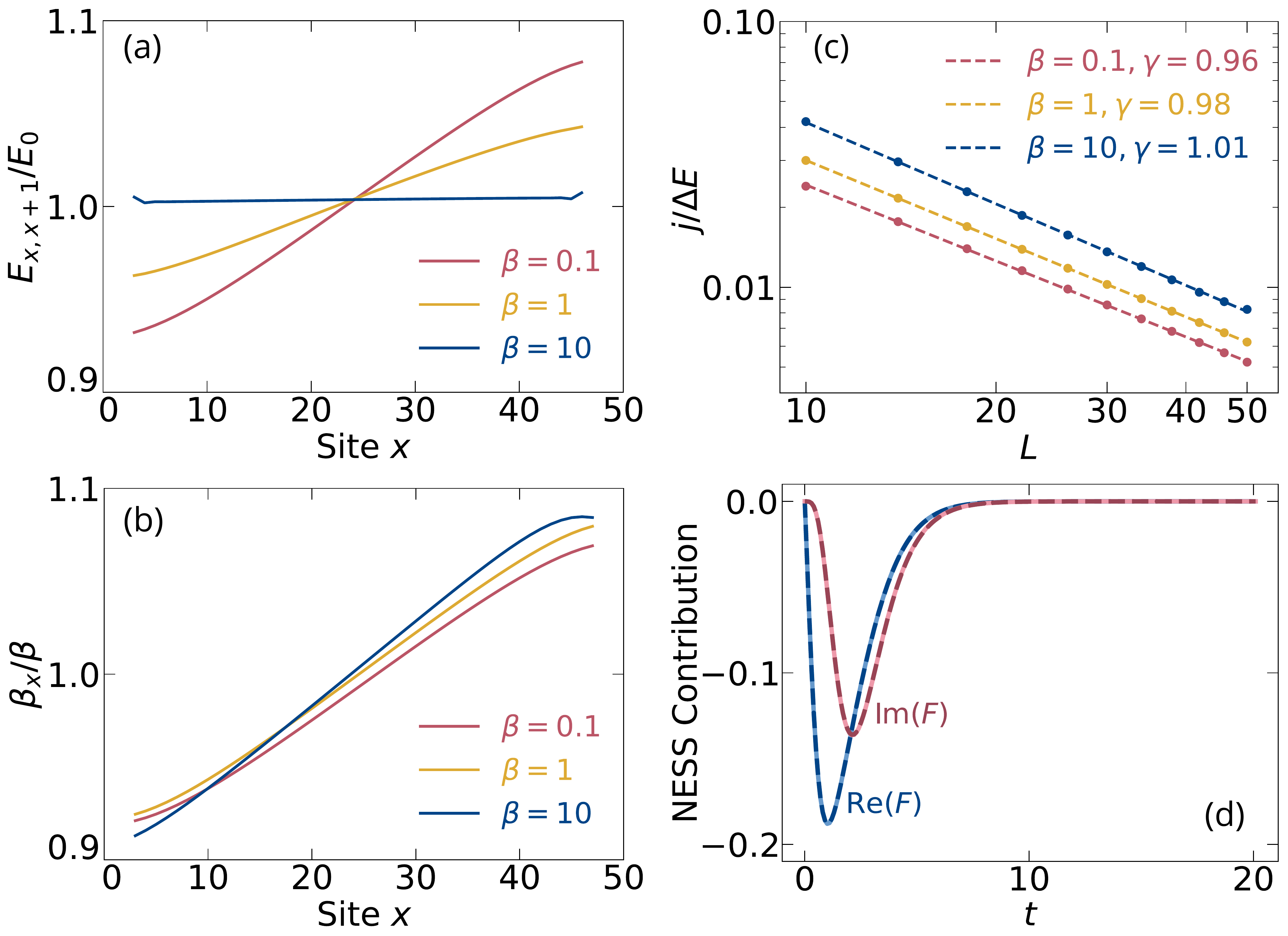}
	\caption{DC NESS transport properties of the boundary-driven SYK chain with $q=4$ at different bath temperatures $\beta$. (a) Spatial energy profiles, rescaled to the equilibrium energy $E_0$ at temperature $\beta$, for a chain of length $L=50$. (b) Same as (a), but for the local inverse temperature. Both energy and temperature profiles are linear in the bulk. (c) Scaled energy current $j/\Delta E$ as a function of system size. Symbols denote numerical values and lines represent fits to the scaling $j/\Delta E = -D/L^\gamma$. The values of $\gamma$ suggest diffusive transport according to Fourier's law. (d) Derivative of the equilibrium Green's function $\diff G_0(t)/\diff\beta$ (dashed lines) showing agreement with the non-equilibrium contribution $F(t)$ (solid lines) to the Green's functions in \eqref{eq:dc_ness_ansatz}, as extracted numerically from NESS. Both are scaled to unit norm.}
	\label{fig:fig3}
	\end{center}
\end{figure}

In the case of boundary driving, we introduce a small temperature imbalance $\delta\beta$ at the two ends of the chain. \figref{fig:fig3} showcases our findings for $q=4$, but similar results hold for all values of $q$. We should mention that although the SYK model with $q=2$ corresponds to free fermions, is non-integrable, and does not satisfy ETH~\cite{garcia2018,haque2019,bhattacharya2019}, it still slowly thermalizes when coupled to external baths and conducts energy in the same way as its analogs with higher $q$. In \figref{fig:fig3}(a) we plot the on-bond energy $E_{x, x+1}$, normalized by its equilibrium value $E_0$, as a function of lattice site $x$. The energy profiles are linear in the bulk, with small boundary effects that become more prominent at low temperatures. Moreover, we find that local Green's functions obey the FDT with inverse temperatures that interpolate linearly between $\beta-\delta\beta$ and $\beta+\delta\beta$ at the two ends, as shown in \figref{fig:fig3}(b). The temperature in the middle of the chain precisely agrees with the average bath temperature. This is different from chaotic spin models, where one usually finds a much higher local temperature in the bulk~\cite{zanoci2021}.
We also find that the current scales linearly with inverse system size at all temperatures, as displayed in \figref{fig:fig3}(c). Notice that the scaling exponent $\gamma$ is close to one, which confirms that our model is a diffusive metal. 

\begin{figure}[htp]
	\begin{center}
	\includegraphics[width = \columnwidth]{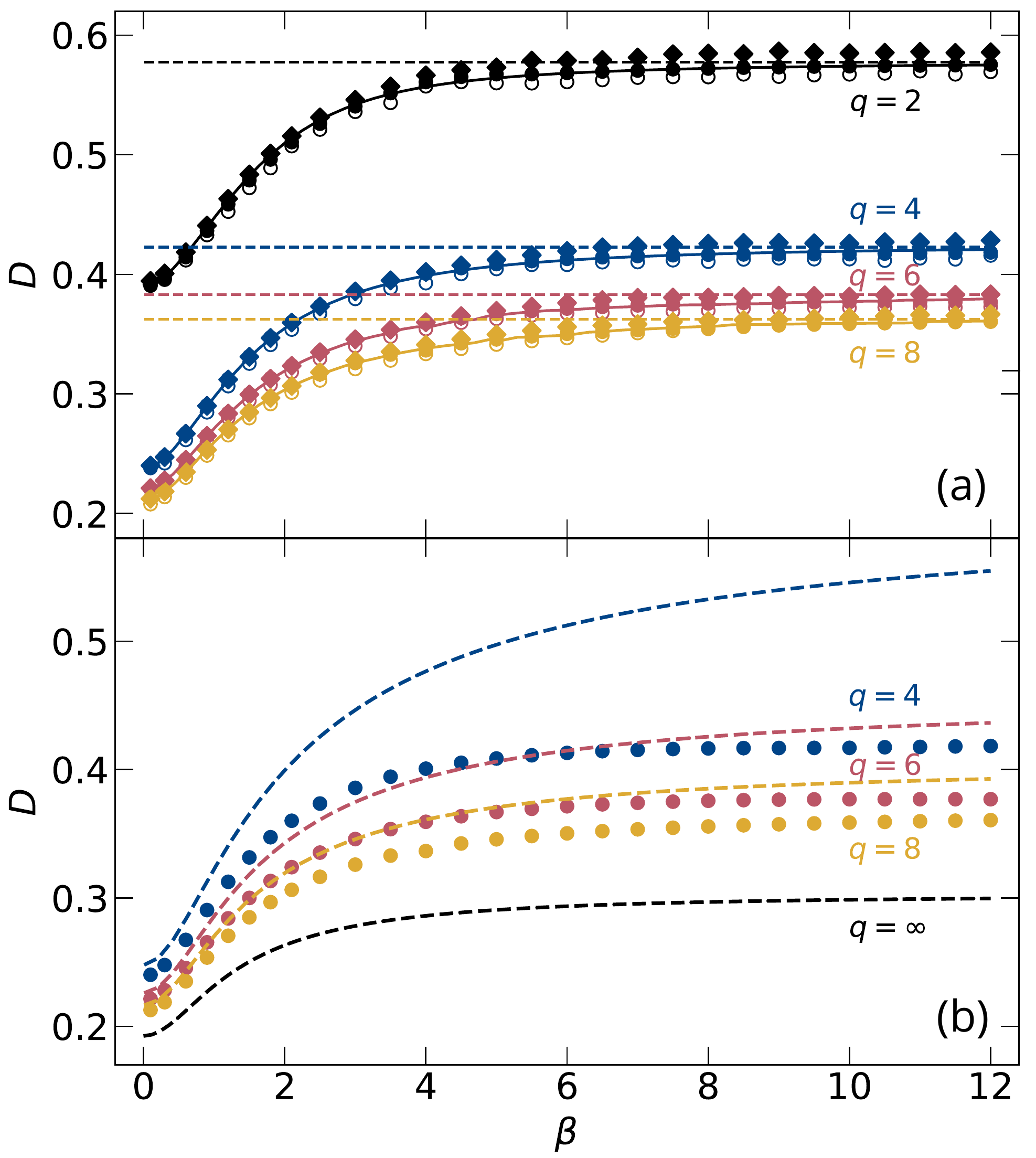}
	\caption{Temperature dependence of the DC diffusion constant for different values of $q$. Filled circles represent numerical values for the boundary-driven SYK chain in both panels. (a) Empty circles (shaded diamonds) correspond to the case of weak (strong) bulk driving. Solid lines represent the diffusion constant computed using the ansatz $F(t) = dG_0(t)/d\beta$, while the dashed lines indicate the conformal limit derived in Ref.~\cite{gu2017diffusion}. (b) Dashed lines show the diffusion constants in the large $q$ limit.}
	\label{fig:fig4}
	\end{center}
\end{figure}

In addition, we observe a very fast convergence of the diffusion constant with system size, which is likely due to the all-to-all nature of the couplings in the SYK model. Therefore, when computing the temperature dependence of the diffusion coefficient, we can restrict our attention to moderately-sized systems with $L=20$ sites. Our results for $D(\beta)$ at various $q$ can be found in \figref{fig:fig4}(a). The transport coefficient increases with $\beta$ and approaches a $q$-dependent constant both in the limit of zero and infinite temperature. The answer for the diffusion constant of an SYK chain in the low-temperature conformal limit has been previously derived in Ref.~\cite{gu2017diffusion} by studying the leading order contributions to the fermionic four-point functions. Their derivation can be generalized to the case of arbitrary $q$ in a straightforward way and we obtain 
\begin{equation}
    D = \frac{q}{q-1}\frac{\pi}{2\alpha_K}\frac{\J_1^2}{\J},
    \label{eq:inf_beta_theory}
\end{equation}
where $\J = J\sqrt{q2^{1-q}}$ is the rescaled effective coupling introduced in \secref{sec:equilibrium} and $\alpha_K$ is a numerical constant whose value is close to $3$ (e.g. Fig. 9 in \cite{maldacena2016}). We recover the original expression in Ref.~\cite{gu2017diffusion} by setting $q=4$. The conformal answer in \eqref{eq:inf_beta_theory} is indicated by dashed lines in \figref{fig:fig4}(a). We observe a striking agreement with our numerical results at large $\beta$.

Most importantly, the authors also showed that in the conformal limit, the energy diffusion constant $D$ is related to the butterfly velocity $v_B$ via the simple relation $D=\beta v_B^2/2\pi$, which realises a conjectured bound on diffusion in incoherent metals~\cite{hartnoll2014,blake2016_1,blake2016_2,blake2017,hartman2017,choi2021}. We further explore the connections between transport and many-body chaos in \appref{sec:appendixB}, where we show that the energy diffusion is upper bounded by chaos $D\leq v_B^2/\lambda_L$, with $\lambda_L$ denoting the Lyapunov exponent. The same conclusion has been reached for the energy diffusion constant in inhomogeneous SYK chains~\cite{gu2017} and the charge diffusion constant in a holographic model~\cite{lucas2016}.

%Finally, the authors of Ref.~\cite{gu2017diffusion} use a different method to obtain the low-temperature diffusion constant of an SYK chain with $q=4$. We can easily generalize their derivation to the case of arbitrary $q$ and compare the result with the diffusion constant obtained above. We will only point out the relevant equations that need to be generalized to arbitrary $q$ and the rest of the derivation follows Ref.~\cite{gu2017diffusion}. First, the spatial kernel is modified to

%\begin{equation}
%    s(p) = 1+\frac{2J_1^2}{3J^2}(\cos(p)-1) \to 1+\frac{qJ_1^2}{2(q-1)J^2}(\cos(p)-1) \approx 1-\frac{qJ_1^2}{2(q-1)J^2}\frac{p^2}{2}.
%\end{equation}
%Next, the $h=2$ eigenvalue becomes (see \cite{maldacena2016})

%\begin{equation}
%    k(h=2, n) = 1-\frac{\sqrt{2}\alpha_K}{\beta J}|n| \to 1-\frac{\alpha_K}{\beta \J}|n|,
%\end{equation}
%where $\alpha_k\approx 3$ in the large $q$ limit (see Fig.9 in \cite{maldacena2016}). Combining these two modified equations, we deduce that

%\begin{equation}
%    1-s(p)k(h=2, n) \approx \frac{qJ_1^2}{4(q-1)J^2}p^2 + \frac{\alpha_K}{\beta \J} = \frac{\alpha_K}{2\pi\J}\Big(\frac{2\pi|n|}{\beta} + \frac{\pi q\JJ_1^2}{2(q-1)\alpha_KJ^2}p^2\Big)\equiv \frac{\alpha_K}{2\pi\J}\Big(\frac{2\pi|n|}{\beta} +Dp^2\Big).
%\end{equation}

\subsection{Bulk-driven SYK chain}
\label{sec:bulk_dc}

Our non-equilibrium driving scheme with baths on the edges can be used in principle to extract the diffusion constant of arbitrary models at any temperature. However, in practice, the times for which we have to numerically evolve the KB equations until convergence can be quite long, especially at low temperatures and for the $q=2$ case. The main computational cost comes from the fact that the influence of the bath has to propagate from the boundary all the way to the middle of the system. Therefore, it seems natural to search for an alternative setup which circumvents this problem.  

Motivated by the linearity of the temperature profile along the chain, we propose to couple each site $x$ to a bath at inverse temperature $\beta_x$ which linearly interpolates between $\beta_L$ and $\beta_R$, as shown in \figref{fig:fig1}(b). We have already studied the equilibrium properties of the system in the presence of this bulk driving in \secref{sec:equilibrium}. We identified two scenarios in which the effective coupling $\J$ in the bulk remains mostly unchanged. In the case of strong bulk driving, we set $\mathcal{V}=\J_0=1/\sqrt{2}$ and in the case of weak bulk driving, we choose $\mathcal{V}=0.2$ and $\J_0=1$. We expect that the bulk driving does not significantly alter the properties of our system even in non-equilibrium and its diffusion coefficient remains unchanged as long as $\J$ and $\J_1$ stay the same in the bulk. Indeed, in the low-temperature limit we can confirm this directly using \eqref{eq:inf_beta_theory}. For the more general case, we solve the KB equations in the presence of bulk driving and plot our results in \figref{fig:fig4}(a). We see that the diffusion constants for both strong and weak driving agree well with the results obtained via boundary driving, with the weak driving showing a slightly better match at low temperatures. 

\begin{figure}[htp]
	\begin{center}
	\includegraphics[width = \columnwidth]{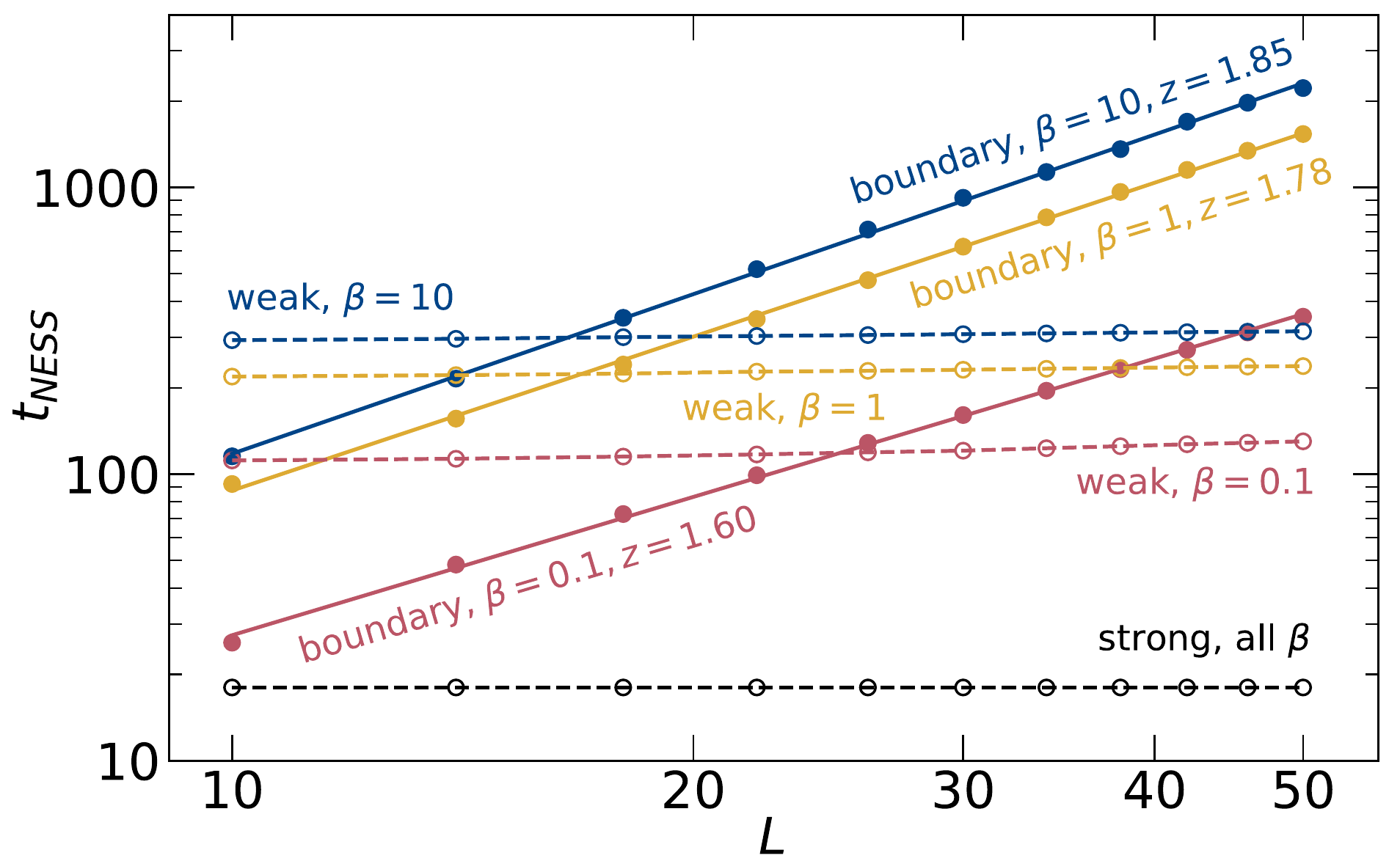}
	\caption{Convergence time of the non-equilibrium dynamics for boundary (solid circles) and bulk (empty circles) driving at various temperatures. The boundary-driven systems exhibit a power law scaling $t_{\mathrm{NESS}}\sim L^z$ with system size (solid lines), while both weak and strong bulk driving lead to nearly-constant convergence times (dashed lines). The convergence threshold was set to $\eta=10^{-3}$.}
	\label{fig:fig5}
	\end{center}
\end{figure}

However, the main appeal of using these alternative driving schemes is their fast convergence times. We compare the times required to reach NESS in \figref{fig:fig5}. For the setup with baths at the boundaries, we find a clear power law dependence of $t_{\mathrm{NESS}}$ on system size. Notice that the convergence time scales superlinearly with $L$ and the exponent is somewhat higher at low temperatures. On the other hand, $t_{\mathrm{NESS}}$ is approximately constant in the case of weak driving and the advantage over boundary driving is already noticeable around $L\approx20$. Furthermore, strong bulk driving leads to an even lower size- and temperature-independent convergence time. These scalings are roughly the same for all $q$. Hence, there is a clear computational advantage in adding baths to the bulk with little to no impact on the transport properties of the system. We expect that a similar speed-up can be achieved in other spin models. 

It is important to emphasize that using bulk driving to exactly match the results of boundary driving was only possible due to the simple, predictable nature of the temperature profile across the system. Moreover, we relied on the fact that the baths are similar to the system SYK clusters (i.e. same $q$, similar $\beta$ etc.). This may no longer be the case in other, more complex non-equilibrium scenarios. In fact, we will briefly discuss this in the context of AC transport in \secref{sec:bulk_ac}.

\subsection{Non-equilibrium ansatz}
\label{sec:ness_ansatz_dc}

Our analysis of the steady-state observables in the previous sections suggests a relatively simple ansatz for the local non-equilibrium Green's function that emerges at late times. In equilibrium, the Green's function deep in the bulk takes on a site-independent and time translation-invariant form $G^{(0)}(t)$, which is the solution to the SD equation in Lorentzian time with an effective coupling $J$ and inverse temperature $\beta$. In the presence of a small uniform bias, we expect the non-equilibrium Green's function to have an extra site-dependent correction proportional to the temperature gradient. Additionally, we expect the Green's function to become time translation-invariant at late times, which we indeed observe in our simulations. Therefore, we look for a NESS solution to the KB equations of the form 
\begin{equation}
    G_x^>(t) = G^{(0)}(t) + xF(t),
    \label{eq:dc_ness_ansatz}
\end{equation}
where $|F(t)|\ll |G^{(0)}(t)|$ is the non-equilibrium contribution. Similarly, we can expand the self-energy as
\begin{equation}
    \Sigma_x^>(t) = -i^qJ^2G^{(0)}(t)^{q-2} \left(G^{(0)}(t)+x(q-1)F(t)\right).
    \label{eq:sigma_ness_ansatz}
\end{equation}

These expressions already lead to a much simpler version of the KB equations in~(\ref{eq:KB}) for the non-equilibrium contribution $F(t)$. However, we can find an explicit solution for $F(t)$ by exploiting the fact that the inverse temperature gradient $\nabla\beta$ is constant in the bulk (see \figref{fig:fig3}(b)). Assuming that the effective coupling $J$ also remains constant, we can perform a first-order expansion in the small gradient
\begin{equation}
    G_x^>(t) \approx G^{(0)}(t) + x\frac{\diff G^{(0)}(t)}{\diff\beta}\nabla\beta,
    \label{eq:dc_ness_expansion}
\end{equation}
This is akin to a gradient expansion in hydrodynamics. By comparing this to \eqref{eq:dc_ness_ansatz}, without loss of generality, we can identify $F(t)\equiv \diff G^{(0)}(t)/\diff\beta$. Notice that the overall magnitude of $F(t)$ does not matter, since to first order, both the energy gradient and the current will be proportional to $F$. This is consistent with the expectation that the exact value of the temperature bias $\delta\beta$ should not affect the transport properties, as long as we are within the linear response regime. A comparison between $F(t)$ extracted numerically as the difference between two consecutive Green's functions in NESS and the derivative $\diff G^{(0)}(t)/\diff\beta$ is presented in \figref{fig:fig3}(d). As we argued above, the two functions coincide. 

Furthermore, one can easily verify using \eqref{eq:dc_ness_ansatz} and \eqref{eq:sigma_ness_ansatz} that if $G^{(0)}(t)$ is a solution to the KB equations in~(\ref{eq:KB}), then so is $G_x^>(t)$ in \eqref{eq:dc_ness_expansion}, to first order in the gradient. We emphasize that $F(t)= \diff G^{(0)}(t)/\diff\beta$ is a solution to the KB equations, but it is not necessarily the unique solution. One could in principle devise more complicated non-equilibrium setups and initial conditions where \eqref{eq:dc_ness_ansatz} holds, but the non-equilibrium correction $F(t)$ has a more complicated form. This is why the numerical comparison in \figref{fig:fig3}(d) is crucial. 

Now we can use this ansatz to obtain a numerical solution for the diffusion constant at arbitrary temperature. To first order in $F$, the energy gradient in \eqref{eq:gradient_1D} becomes 
\begin{equation}
    \nabla E = -2i^{q+2}J^2 \int_0^{\infty} \diff t \im\left[G^{(0)}(t)^{q-1} F(t)\right] = \frac{\diff E_0}{\diff\beta}.
    \label{eq:grad_E}
\end{equation}
where $E_0$ is the equilibrium energy computed from $G^{(0)}(t)$ and we have used the identities in \eqref{eq:gtr_lsr} to write the integral only in terms of functions evaluated at $t\geq0$. Similarly, the current in \eqref{eq:current_1D} becomes
\begin{equation}
    j = \frac{1}{2}J_1^2J^2\Re(j_{++}+j_{+-}),
\end{equation}

\begin{widetext}
\begin{equation}
\begin{split}
    j_{++}+j_{+-} & = -2iq \int_0^{\infty} \diff t \int_t^\infty \diff t' G^{(0)}(t')^{q-2} G^{(0)}(t'-t) \left(G^{(0)}(t')\im\left[F(t)G^{(0)}(t)^{q-2}\right] -F(t')\im\left[G^{(0)}(t)^{q-1}\right]\right) \\
    &= -\frac{2iq}{q-1} \int_0^{\infty} \diff t \int_t^\infty \diff t' G^{(0)}(t')^{2q-2} G^{(0)}(t'-t) \frac{\diff}{\diff\beta}\left(\frac{\im\left[G^{(0)}(t)^{q-1}\right]}{G^{(0)}(t')^{q-1}}\right).
    \label{eq:j}
\end{split}
\end{equation}
\end{widetext}
Note that both of these quantities are independent of $x$ and we can compute the diffusion constant $D=-j/\nabla E$ solely in terms of the equilibrium function $G^{(0)}(t)$.

In general, the Green's function does not have a closed-form representation, except for the case of free fermions $q=2$, which we discuss in \appref{sec:appendixC}. Therefore, we have to numerically solve the SD equation to get $G^{(0)}(t)$ ~\cite{eberlein2017,guo2019}, then perform the integrals above to obtain $D$. The final result is shown as the solid lines in \figref{fig:fig4}(a) and it agrees perfectly with diffusion constants extracted from solving the full KB equations in the presence of baths (solid circles in \figref{fig:fig4}(a)). 

Our non-equilibrium ansatz significantly reduces the complexity of computing the transport properties of the SYK chain compared to the boundary- or bulk-driven setups. Moreover, under certain circumstances, such as the large $q$ limit discussed below, we can use this ansatz to find simple expressions for the diffusion constant.

%Next, we plug this ansatz into the KB equations. We will focus on \eqref{eq:KB_1} for now. Using \eqref{eq:gtr_lsr} and expanding to first order in $F$, we get

%\begin{align*}
%    i\partial_{t_1}\big(G^{(0)}(t_1-t_2) &+ xF(t_1-t_2)\big) =-i^qJ^2\int_{-\infty}^{t_1}dt' \bigg(G^{(0)}(t_1-t')^{q-1} + x(q-1)G^{(0)}(t_1-t')^{q-2}F(t_1-t')\\
%    &+G^{(0)}(t'-t_1)^{q-1} + x(q-1)G^{(0)}(t'-t_1)^{q-2}F(t'-t_1) \bigg)\bigg( G^{(0)}(t'-t_2) + xF(t'-t_2)\bigg) \\
%    &+ i^qJ^2\int_{-\infty}^{t_2}dt' \bigg(G^{(0)}(t'-t_2) + xF(t'-t_2)+G^{(0)}(t_2-t') + xF(t_2-t') \bigg) \\
%    &\cdot \bigg( G^{(0)}(t_1-t')^{q-1} + x(q-1)G^{(0)}(t_1-t')^{q-2}F(t_1-t')\bigg) \\
%    &= -2i^{q+1}J^2\int_{-\infty}^{t_1}dt' \im\bigg[G^{(0)}(t'-t_1)^{q-1} + x(q-1)G^{(0)}(t'-t_1)^{q-2}F(t'-t_1) \bigg]\\
%    &\cdot \bigg( G^{(0)}(t'-t_2) + xF(t'-t_2)\bigg) \\
%    &+ 2i^{q+1}J^2\int_{-\infty}^{t_2}dt' \im\bigg[G^{(0)}(t_2-t') + xF(t_2-t') \bigg] \\
%    &\cdot \bigg( G^{(0)}(t_1-t')^{q-1} + x(q-1)G^{(0)}(t_1-t')^{q-2}F(t_1-t')\bigg).
%\end{align*}

%Recall that the equilibrium solution $G^{(0)}$ also satisfies a KB equation, so we can cancel the equilibrium contribution

%\begin{align*}
%    &i\partial_{t_1}F(t_1-t_2) =\\
%    &-2i^{q+1}J^2\int_{-\infty}^{t_1}dt'\bigg((q-1)G^{(0)}(t'-t_2) \im\big[G^{(0)}(t'-t_1)^{q-2}F(t'-t_1)\big] +F(t'-t_2) \im\big[G^{(0)}(t'-t_1)^{q-1}\big]\bigg) \\
%    &+ 2i^{q+1}J^2\int_{-\infty}^{t_2}dt' \bigg(G^{(0)}(t_1-t')^{q-1} \im\big[F(t_2-t')\big] +(q-1)G^{(0)}(t_1-t')^{q-2}F(t_1-t') \im\big[G^{(0)}(t_2-t')\big]\bigg).
%\end{align*}
%Note that the $x$ dependence also disappears.

\subsection{Large \texorpdfstring{$q$}{q} limit}
\label{sec:large_q_results}

We now apply our construction in the large $q$ limit introduced in \secref{sec:large_q}. For a 1D chain, \eqref{eq:KB_large_q_trans_inv} away from the boundary can be written as 

\begin{equation}
    -\frac{\partial^2 g_x(t)}{\partial t^2} =  2\J_0^2 e^{g_x(t)} + 2\J_1^2 e^{\frac{g_x(t)}{2}}\left(e^{\frac{g_{x-1}(t)}{2}}+e^{\frac{g_{x+1}(t)}{2}}\right).
\end{equation}
In equilibrium, $g^{(0)}(t)$ derived in \eqref{eq:g_0} is a uniform solution. However, in the presence of a small temperature gradient, we look for a general solution of the form 

\begin{equation}
    g_x(t) = g^{(0)}(t) + xf(t),
\end{equation}
where $|f(t)|\ll |g^{(0)}(t)|$ is the non-equilibrium correction. Just as before, we identify $f(t)\equiv \diff g^{(0)}(t)/\diff\beta$. In terms of this new function we have $F(t)=G^{(0)}(t)f(t)/q$. Since $f$ is proportional to the imposed temperature gradient, we can make it arbitrarily small and only keep terms linear in $f$. The KB equation becomes

\begin{equation}
    \frac{\partial^2 f(t)}{\partial t^2} = -2\J^2 e^{g^{(0)}(t)} f(t), 
    \label{eq:ODE}
\end{equation}
where $\J^2 = \J_0^2 + 2\J_1^2$ and we canceled the equilibrium terms. We supplement this ODE with the initial condition $f(0) = 0$, imposed by $g_x(0) = g^{(0)}(0)=0$. However, since the equation is of second order, we have to also specify $f'(0)$, on which we have (almost) no restrictions. This is a consequence of the aforementioned freedom in choosing the overall scale of $f$. Note that $f(t)$ also has to obey the identity in \eqref{eq:large_q_sym}.  

Given our guess for $f(t)$ and \eqref{eq:g_0}, we find the following non-equilibrium contribution

\begin{equation}
    f(t) = -\tan(\frac{\pi v}{2}) + \Big(1+i\J t\sin(\frac{\pi v}{2})\Big)\tan(\frac{\pi v}{2} - \frac{i\pi vt}{\beta})
    \label{eq:f_large_q}
\end{equation}
It is easy to check that this is indeed a solution to \eqref{eq:ODE} and that it satisfies all the conditions mentioned above. In the large $q$ limit, the energy gradient reads

\begin{equation}
\begin{split}
    \nabla E &= \frac{\J^2}{q^2}\int_0^{\infty} \diff t \im\left[e^{g^{(0)}(t)} f(t)\right] \\
    &= -\frac{1}{2q^2}\int_0^{\infty} \diff t \im\left[\frac{\partial^2 f(t)}{\partial t^2}\right]\\
    &= \frac{1}{2q^2}  \im\left[f'(0) - f'(\infty)\right] \\
    &= -\frac{\J}{2q^2}\cos(\frac{\pi v}{2}),
    \label{eq:grad_E_1}
\end{split}
\end{equation}
where we used \eqref{eq:ODE} in the second line. The formula for the current is a bit more involved

\begin{widetext}
\begin{equation}
\begin{split}
    j &= \frac{\J_1^2\J^2}{2q^2}\int_{0}^{\infty} \diff t \int_{t}^{\infty} \diff t' \Bigg[\re\left[e^{\left(1-\frac{1}{q}\right)g^{(0)}(t)}f(t)\right]\im\left[e^{\left(1-\frac{1}{q}\right)g^{(0)}(t')+g^{(0)}(t'-t)/q}\right] \\
    &- \re\left[e^{\left(1-\frac{1}{q}\right)g^{(0)}(t)}\right]\im\left[e^{\left(1-\frac{1}{q}\right)g^{(0)}(t')+g^{(0)}(t'-t)/q}f(t')\right] \Bigg].
    \label{eq:j_general}
\end{split}
\end{equation}
\end{widetext}

Unfortunately, this double integral does not have a closed-form solution for arbitrary $q$ and $\beta$. However, we are able to approximate it in both the $q\to\infty$ limit with arbitrary $\beta$ and in the $\beta\to0$ and $\beta\to\infty$ limits with arbitrary $q$. The details of this computation are provided in \appref{sec:appendixC}. For example, the $q\to\infty$ limit yields 

\begin{equation}
    D = \frac{\J_1^2}{3\J}\left(\frac{\pi v}{2}\sin(\frac{\pi v}{2})+\cos(\frac{\pi v}{2})\right).
\end{equation}
The temperature dependence enters this expression implicitly through $v$ (see \eqref{eq:v}). Note that this equation exactly matches the one derived in Ref.~\cite{choi2021} from the energy density two-point function, thus providing an independent consistency check for our NESS calculation. 

Our non-equilibrium ansatz led to a remarkably simple formula for the diffusion constant in the limit of infinite $q$. We plot this result in \figref{fig:fig4}(b). In the same figure, we also show the diffusion constants obtained by numerically integrating \eqref{eq:j_general}. The large $q$ expansion curves follow the same trend as the previously obtained exact results. The agreement between the two is significantly better at higher temperatures and larger $q$. The diffusion constant clearly decreases with $q$ and we expect it to eventually approach the $q=\infty$ result (black line in \figref{fig:fig4}(b)).

\subsection{Models with \texorpdfstring{$q_S\neq q_I$}{different q}}
\label{sec:other_models}

\begin{figure}[htp]
	\begin{center}
	\includegraphics[width = \columnwidth]{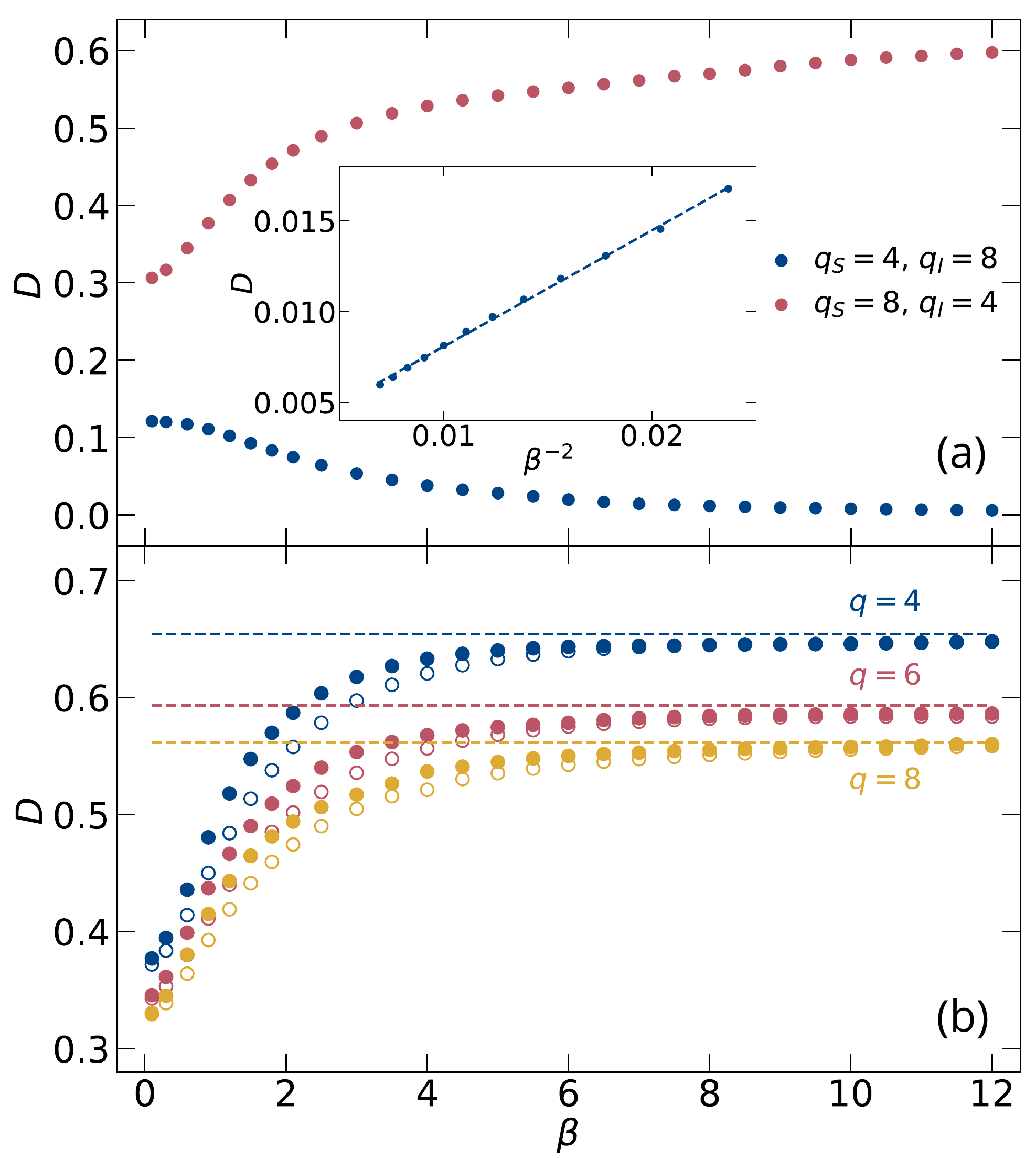}
	\caption{Temperature dependence of the diffusion constant for other SYK systems. (a) One-dimensional chain with $q_S > q_I$ (red) and $q_S < q_I$ (blue). For the latter, the inset shows a quadratic fit $D\sim \beta^{-2}$ in the conformal limit. (b) Two-dimensional uniform square lattice. Filled circles represent numerical values obtained from the KB equations, while empty circles denote values inferred from one-dimensional data. Dashed lines indicate the conformal limit in \eqref{eq:2D_conformal}.}
	\label{fig:fig6}
	\end{center}
\end{figure}

In all the examples previously studied in this paper, we found that our systems behaved like diffusive metals with a finite energy diffusion constant at both zero and infinite temperatures. However, this is not the case for all SYK chains. In this section, we will discuss a family of models with $q_S\neq q_I$, which were first introduced in the context of metal to insulator transitions~\cite{cmjian2017}. First, we consider a model with $q_I=2q_S=2q$, where $q\geq2$. This model is equivalent to isolated SYK clusters with $q_S=q$ in the IR limit, since the inter-cluster coupling becomes irrelevant. Hence, the system should become an insulator at low temperatures~\cite{cmjian2017}. To confirm this prediction, we compute the diffusion constant as a function of temperature using the boundary-driven setup for $q=4$. The results are shown in \figref{fig:fig6}(a). We can see that the diffusion constant approaches zero in the low-temperature limit. In fact, we can further observe that $D$ decreases quadratically with temperature, as depicted in the inset. This agrees with the large $q$ calculation of Ref.~\cite{cmjian2017}, which predicts a decay $D\sim\beta^{-2}$.  

Next, we consider the opposite model where $q_S=2q_I=2q$. In this model, we expect the on-site coupling $\J_0$ to become irrelevant in the IR, where the low-temperature transport should be governed by the intra-cluster coupling $\J_1$. The system is expected to behave like a diffusive metal, which we confirm numerically in \figref{fig:fig6}(a) for the case $q=4$. The diffusion constant approaches a constant both at the high and low temperatures. According to the large $q$ analysis of Ref.~\cite{cmjian2017}, the diffusion constant should converge to 
\begin{equation}
    D = \frac{\pi\J_1}{3\sqrt{1+\J_0^2/8\J_1^2}}\approx0.99,
\end{equation}
for $\J_0=\J_1=1$. Although our results seem to converge to a somewhat smaller value, this discrepancy is on par with the one in our other large $q$ predictions in \figref{fig:fig4}(b). 

\subsection{Higher-dimensional generalizations}
\label{sec:higher_dim}

So far, we have focused on one-dimensional SYK chains, but our analysis can be easily generalized to higher-dimensional lattices. For example, we can consider a two-dimensional $L_x\times L_y$ square lattice with baths attached along the vertical boundaries, as shown in \figref{fig:fig1}(c). Let the lattice dimensions be $L_x=15$ and $L_y=4$, with periodic boundary conditions in the $y$-direction. The boundary driving will impose a current $j_x$ in the $x$-direction, but there will be no net current in the $y$-direction due to symmetry. We time-evolve the KB equations in the usual way and use \eqref{eq:current_final} to compute the current by summing contributions from all four neighbors of a given site. 

Our results for the square lattice are given in \figref{fig:fig6}(b). We observe the same temperature dependence as in the one-dimensional case. The asymptotic value at low temperatures can be compared again to the conformal limit answer in Ref.~\cite{gu2017diffusion}. For a translation invariant lattice, we expect the pole determining the diffusion constant to get equal contributions from both the $p_x^2$ and $p_y^2$ momenta, thus doubling the value in \eqref{eq:inf_beta_theory}

\begin{equation}
    D = \frac{q}{q-1}\frac{\pi}{\alpha_K}\frac{\J_1^2}{\J},
    \label{eq:2D_conformal}
\end{equation}
where $\J=\sqrt{\J_0^2+4\J_1^2}$ is now the effective coupling in 2D. This conformal answer is depicted by dashed lines in \figref{fig:fig6}(b) and it agrees well with our data. 

Given the simplicity of the low-temperature answer above, one could attempt to estimate the two-dimensional diffusion constant at arbitrary temperatures by appropriately rescaling the one-dimensional data

\begin{equation}
    D_{\mathrm{2D}} = 2\sqrt{\frac{\J_0^2+2\J_1^2}{\J_0^2+4\J_1^2}}D_{\mathrm{1D}},
\end{equation}
where we took into account the difference in effective coupling. This guess is represented by empty circles in \figref{fig:fig6}(b) and it is in surprisingly good agreement with the actual 2D data, even though the diffusion constant at high temperatures has a more complicated dependence on $\J$. This result suggests that diffusion in higher dimensional SYK lattices is very similar to its one-dimensional counterpart. 

%Further generalizations to $d$-dimensional hypercubic lattices are straightforward. 

\section{AC Transport}
\label{sec:ac}

We turn our attention to the frequency dependence of the diffusion coefficient $D(\omega)$. This quantity is fundamental, as it describes the response of the system to an external time-dependent bias with angular frequency $\omega$. The frequency dependence is typically computed via the Kubo formula~\cite{kubo1957}, which relates transport coefficients to current autocorrelation functions. However, our framework is more suitable for directly measuring the current that emerges as a result of periodic driving.

In order to induce a time-dependent energy gradient in our system, we must couple it to baths at inverse temperatures $\beta_{R,L}=\beta\pm\delta\beta\cos(\omega t)$. Since it is non-trivial to treat a time-dependent temperature directly within the Schwinger-Keldysh formalism, we choose instead to oscillate the on-site coupling of the baths $J_B\pm\delta J_B\cos(\omega t)$, where $\delta J_B\ll J_B$. Note that the KB equations remain virtually unchanged in the presence of a time-dependent coupling~\cite{eberlein2017,bhattacharya2019,kuhlenkamp2020}, which only changes the local energy scale for the baths and leads to an oscillating temperature. In the limit of $\omega\to0$ we exactly recover the temperature imbalance $\beta\pm\delta\beta$ imposed in the DC case.

\begin{figure}[htp]
	\begin{center}
	\includegraphics[width = \columnwidth]{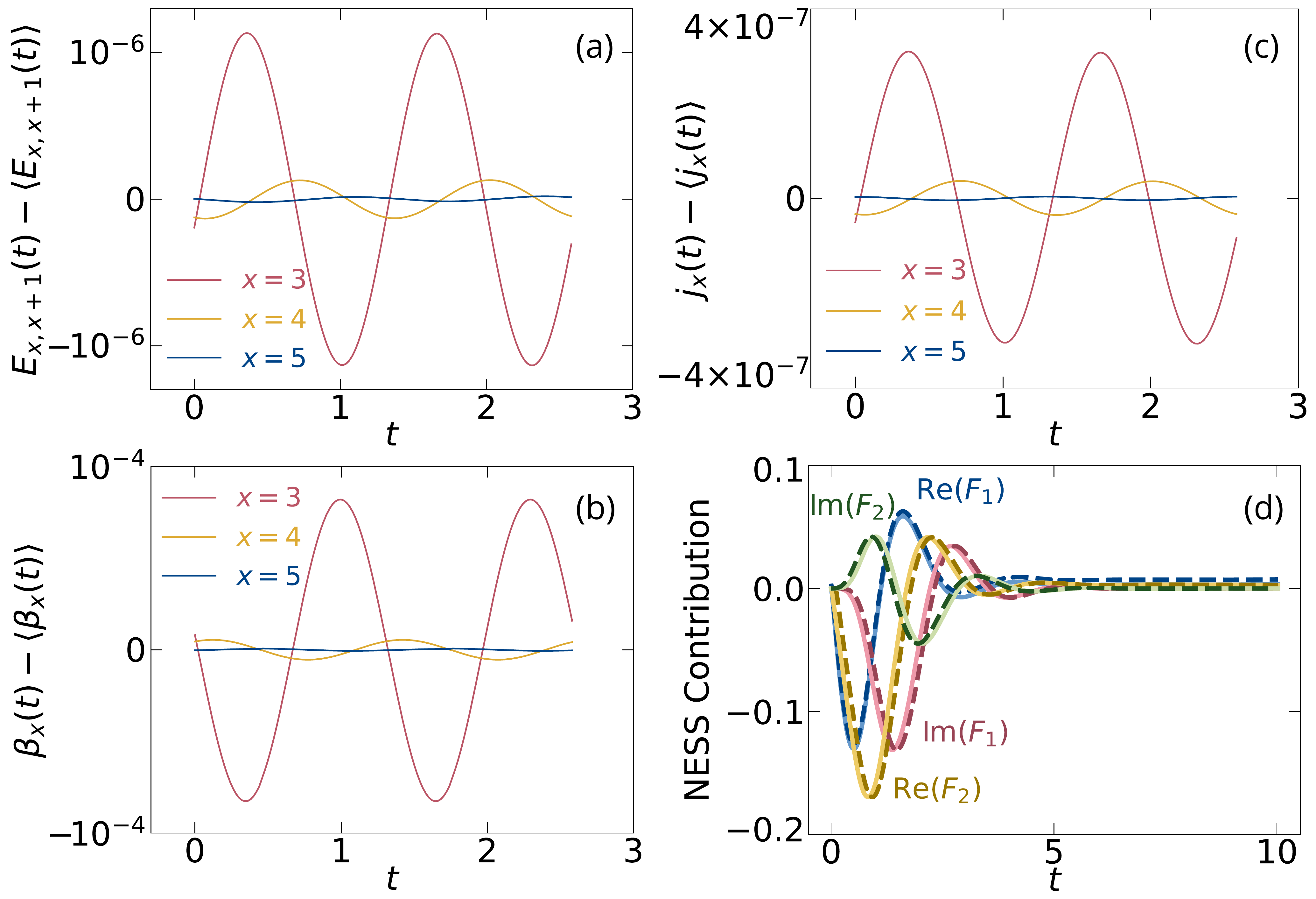}
	\caption{AC NESS transport properties of a boundary-driven SYK chain of length $L=11$ with $q=4$ at $\beta=1$ and $\omega=5$. Time-dependent (a) energy, (b) local inverse temperature, and (c) current for consecutive sites $x$, showing a uniform amplitude decay $c=0.13$ and phase shift $\delta=1.74$. The means were subtracted for easier visualization. (d) The non-equilibrium contributions $F_{1,2}(t)$ to the Green's functions extracted directly from the boundary-driven NESS (solid lines) and computed using the ansatz in \Appref{sec:appendixD} (dashed lines).}
	\label{fig:fig7}
	\end{center}
\end{figure}

We proceed in a manner similar to the DC case. First, we attach the baths at the boundary of our SYK chain and study the resulting NESS. We find that all observables oscillate at frequency $\omega$ with an amplitude that becomes more attenuated the further we measure into the bulk. Based on these insights, we propose a new ansatz for the non-equilibrium Green's function and solve the KB equations numerically using this functional form. We show that the ansatz describes well the AC transport in SYK models. Last, we briefly comment on the setup in the presence of bulk driving. Just as in the case of DC transport, we set $q_S=q_I=q_B=q$ and $\J_0=\J_1=1$.

\subsection{Boundary-driven SYK chain}
\label{sec:boundary_ac}

\begin{figure}[htp]
	\begin{center}
	\includegraphics[width = \columnwidth]{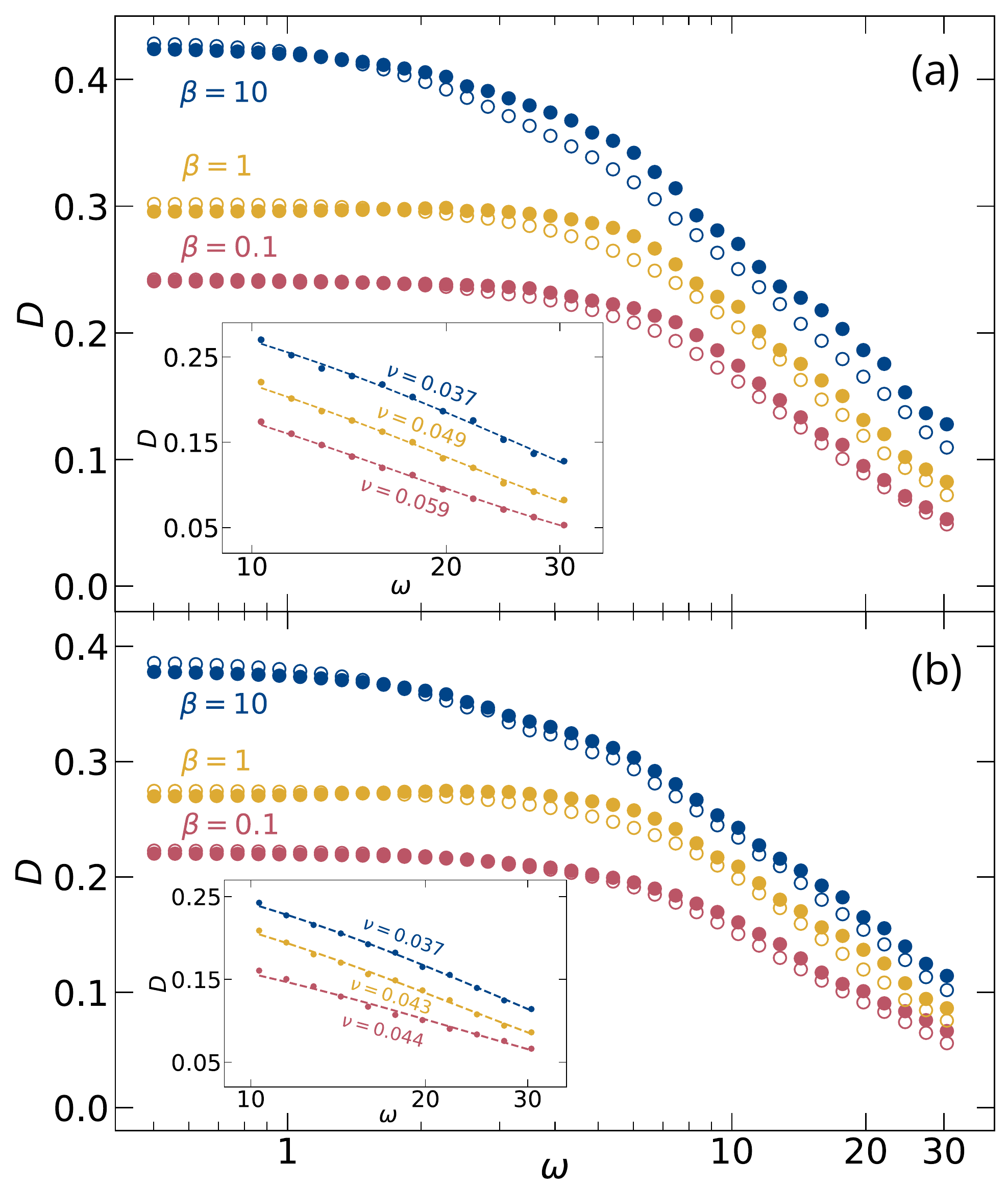}
	\caption{Frequency dependence of the AC diffusion constant for (a) $q=4$ and (b) $q=6$ at various temperatures. Filled circles represent numerical values for the boundary-driven SYK chain. Empty circles correspond to the diffusion constant computed using the ansatz in \eqref{eq:ac_ness_ansatz}. The insets show an exponential scaling $D\sim e^{-\nu\omega}$ at high frequencies.}
	\label{fig:fig8}
	\end{center}
\end{figure}

In the boundary-driven case, the inverse temperatures of the two baths at the ends oscillate out of phase by a half-period, thus always creating an energy gradient and current, whose direction changes in time. A similar configuration has been used to study magnetization transport in a quantum spin chain subject to an oscillating magnetic field bias at the boundaries~\cite{trauzettel2008}. Our results for a sample chain driven at $\omega=5$ are shown in \figref{fig:fig7} and similar outcomes are seen for all regimes of parameters. All local observables experience an oscillatory time-dependence. In \figref{fig:fig7}(a-c) we plot the oscillations of the bond energy, inverse temperature, and current around their equilibrium values for three consecutive sites in the bulk. An interesting feature is that the amplitude of these oscillations is small and decays at a steady rate $c$ as we approach the middle of the chain from either end. Moreover, we observe a constant phase shift $\delta$ from site to site. This indicates an attenuation in the bulk of the wave modes created at the boundary. 

Notice that the current and energy gradient oscillate approximately in phase at every site. Therefore, we can use the ratio of their amplitudes, which is independent of the dampening $c$, to define a diffusion constant $D(\omega) = |j(\omega)/\nabla E(\omega)|$ at each site. Even though the fast decay prevents us from numerically studying large systems, we checked that the diffusion constant converges rapidly with system size and is uniform throughout the chain. We plot our results for $D(\omega)$ at different temperatures in \figref{fig:fig8}. Note that in the zero-frequency limit, we recover the DC values of the diffusion constants from \figref{fig:fig4}(a). Our simulations are limited by the time discretization step at large frequencies, where we have to keep $\diff t$ small to avoid sub-sampling errors, while also evolving the KB equations long enough to achieve convergence. Nonetheless, even in the limited frequency range attainable numerically we observe a clear exponential decay $D\sim e^{-\nu\omega}$ at $\omega\gg J, \beta^{-1}$. The exponent $\nu$ is small and decreases with $\beta$. Moreover, we notice that $\nu$ approaches a $q$-independent value at low temperatures, suggesting a possible universal scaling in the IR.

The exponential decay of the diffusion constant at large frequencies can be explained using recent developments in the theory of periodically-driven many-body systems~\cite{abanin2015,abanin2017,mori2016,kuwahara2016}. Intuitively, the energy exchange between the system and the bath (or its neighbors) becomes inefficient at high frequencies, since it requires significant changes in the many-body state in a very short time~\cite{kuhlenkamp2020}. The amount of energy outputted by the bath during periodic driving is on the order of $\omega$, while the energy that can be absorbed by a local degree of freedom (fermion) is of order $J$. At high frequencies, many fermions (roughly $\omega/J$) would have to work together to absorb this energy over one period. However, such a many-body effect is exponentially suppressed for a Hamiltonian with few-body interactions~\cite{kuwahara2016}. Therefore, the rate of energy exchange between clusters becomes exponentially small, which in turn leads to an exponentially small diffusion coefficient.

\subsection{Non-equilibrium ansatz}
\label{sec:ness_ansatz_ac}

Similarly to the DC case, we use our observations about the NESS to devise an ansatz for the local non-equilibrium Green's function under periodic driving. For convenience, we perform a change of variables from $(t_1, t_2)$ to $t=t_1-t_2$ and $T=(t_1+t_2)/2$. At late times, the Green's functions are no longer time-translation invariant and depend on $T$ in addition to $t$. However, we find that $T$ only enters the oscillatory component of the Green's function. The most general ansatz consistent with these observations and the symmetries of the Green's function (see \eqref{eq:gtr_lsr}) is given by

\begin{equation}
\begin{split}
    G_x^>(t, T) &= G^{(0)}(t) + c^xF(t)\cos(\omega T - x\delta+\phi(t))\\
    &\hspace{-1.45cm} \equiv G^{(0)}(t) + c^x\left(F_1(t)\cos(\omega T -x\delta) + F_2(t)\sin(\omega T -x\delta)\right),
    \label{eq:ac_ness_ansatz}
\end{split}
\end{equation}
where we allowed for a time-dependent phase $\phi(t)$ in addition to the relative phase $\delta$ between sites. The parameter $c$ controls the ratio of amplitudes between neighboring sites. We find it easier to work directly with the sine and cosine components $F_{1, 2}(t)$, rather than the phase $\phi(t)$. In the linear response regime, we have $|F_{1,2}(t)|\ll|G^{(0)}(t)|$. 

We can use this guess to write a set of KB equations for the non-equilibrium functions $F_{1, 2}(t)$. However, the additional factors stemming from the AC driving prevent us from finding an explicit solution in terms of the equilibrium Green's functions. Nevertheless, we can solve the system of equations iteratively to find both $F_{1, 2}(t)$ and $(c, \delta)$, as detailed in \appref{sec:appendixD}. A comparison between $F_{1, 2}(t)$ extracted directly from the NESS Green's functions and those computed numerically in \appref{sec:appendixD} is shown in \figref{fig:fig7}(d). We see a very good agreement between the two. Furthermore, we can use the ansatz to compute $D(\omega)$ at various temperatures. We observe remarkable agreement with the previous data in \figref{fig:fig8}. The slight deviations at high frequencies are due to the extremely small amplitudes of the current and energy gradient in the case of boundary driving. 

\subsection{Bulk-driven SYK chain}
\label{sec:bulk_ac}

Most of the difficulties related to the computation of $D(\omega)$ in the previous two sections arise from the dampening of the oscillations in the bulk. In particular, this has led us to introduce two additional parameters, $c$ and $\delta$, in our non-equilibrium ansatz, which further complicated the calculations. Therefore, it seems natural to consider adding baths in the bulk. Indeed, this would lead to more uniform observables across the chain, corresponding to the limit $c\to1$ and $\delta\to0$. Note that a small, but finite $\delta$ is still required to have non-zero current and can be imposed by staggering the baths. However, a new problem emerges at large frequencies. In this regime, there will be a significant amplitude decay and phase shift between the oscillations in the bath and those in the system at every site. Since the KB equations in the bulk now involve the bath's Green's function, we would have to take into account both the NESS contributions from the system $F_S(t)$, as well as those from the bath $F_B(t)$ due to the time-dependent temperature. The two functions would be related by a new set of parameters $(c', \delta')$ that need to be determined numerically. Therefore, the attenuation problem persists even in the case of bulk driving and we do not get a clear advantage in this case. It is worth noting that in the case of time-independent bulk driving, the bath at fixed temperature did not have a NESS correction $F(t)$ to its Green's function, and hence its influence could be fully captured through renormalizing the effective coupling $J$. 

\section{Discussion}
\label{sec:discussion}

In this paper, we discussed a framework for studying the non-equilibrium properties of a generalized SYK model on arbitrary graphs coupled to thermal baths. We then used this formalism to investigate energy transport in SYK lattices with various $q$-body interactions. Our main focus was on one-dimensional chains, although we also discussed generalizations to higher dimensions. We showed that the transport coefficients can be equivalently computed either by applying a small bias at the boundaries of the system, or by attaching baths to every site in the bulk. Using this setup, we verified that all the models feature diffusive transport, and numerically computed the temperature dependence of the diffusion constants. At low temperatures, we were able to show that the diffusivity approaches a constant value which exactly matches the conformal limit prediction~\cite{gu2017diffusion}. We also showed that energy diffusion is upper bounded by chaos at all temperatures $D\leq v_B^2/\lambda_L$, with equality holding in the conformal regime. It remains to be seen whether this inequality holds for other generalizations of the SYK model.

Our analysis of the non-equilibrium Green's functions that emerge during transport has revealed that they are only weakly perturbed from their equilibrium values. Moreover, we were able to identify the exact functional form of this perturbation for both frequency-independent and frequency-dependent transport. In the DC case, we were able to derive a series of analytical results in the large $q$ expansion of the SYK model. In particular, we obtained a closed-form expression for the diffusion constant at all temperatures in the $q\to\infty$ limit. In the AC case, we managed to solve for the non-equilibrium contributions numerically and showed that the diffusion constant decays exponentially at high frequencies. In both cases, our non-equilibrium ansatz allowed for a very efficient computation of the transport properties of an SYK chain without referencing the specific setup driving the system out of equilibrium. This result is not surprising, since we know that the transport coefficients should not depend on the details of the drive, as long as we are in the linear response regime. Note that although we focused on linear-response-like scenarios in this work, nothing in the general formalism requires this, and it would be interesting to investigate the non-linear response of such SYK networks due to strong driving.

The study of non-equilibrium SYK chains uncovered important insights about the structure of NESS, which can be further applied to transport in more conventional spin systems~\cite{bertini2021,weimer2021,landi2021,zanoci2021}. First, we showed that adding baths in the bulk dramatically improved the convergence time of our simulations. This could be a key ingredient necessary to reach lower temperatures in certain spin systems, where convergence time becomes the limiting factor~\cite{zanoci2021}. Second, we argued that the NESS can be described locally using the non-equilibrium contribution $F$, which acts as a small perturbation on top of the equilibrium solution. A similar approach could be applicable to spin systems where the state is represented as a tensor network~\cite{zanoci2021,bertini2021}. One would proceed by expanding the master equation to first order in $F$ and then solving for the non-equilibrium contribution, in the spirit of \appref{sec:appendixD}. Finally, it would be interesting to study AC transport in spin chains using a periodic drive, similar to the one described in \secref{sec:ac}, and compare the results to those obtained via the Kubo formula~\cite{trauzettel2008,kubo1957,bertini2021}. 

One natural extension of our analysis involves thermoelectric transport in complex SYK models~\cite{sachdev2015,davison2017,song2017,guo2019,cheipesh2020}. This variant of the SYK Hamiltonian is written in terms of complex fermions and features a conserved fermion number in addition to energy~\cite{sachdev2015,davison2017}. Diffusive charge and energy transport has been observed in the strongly correlated metals built from complex SYK clusters~\cite{davison2017,song2017}. They also have other interesting properties, such as a linear in temperature resistivity reminiscent of high-$T_c$ cuprates~\cite{song2017}. It would be instructive to study mixed thermal and electrical transport in these models by explicitly coupling them to baths in various ways.

Given the interesting physical properties of the SYK model and its extensions, multiple experimental realizations~\cite{danshita2017,pikulin2017,chew2017,chen2018} and quantum simulations~\cite{garcia2017,luo2019} of SYK have been proposed. However, there are a few limitations that experiments must overcome. First, all-to-all interactions and the large $N$ limit, which are crucial for exactly solving the SYK model, are hard to achieve in real materials. Second, depending on the Hamiltonian under study, one has to suppress the free-fermion hopping terms, while enhancing the higher-order $q$-body random interactions between the fermions~\cite{franz2018,rahmani2019}. To circumvent these issues, experimental setups in solid-state systems that rely on approximate symmetries to forbid two-body interactions, but allow higher-order couplings, have been proposed. For example, implementations with Majorana modes on the surface of a topological insulator~\cite{pikulin2017} and semiconductor quantum wires coupled to a quantum dot~\cite{chew2017} have been discussed. Additionally, graphene flakes with irregular boundaries in the presence of strong magnetic fields have been suggested as realizations of the complex SYK model~\cite{chen2018,kuhlenkamp2020}. These experiments provide a promising path towards studying out-of-equilibrium dynamics in SYK models. Actual measurements of the diffusion constants could potentially be performed on these devices in the near future. 

\begin{acknowledgments}
We would like to thank Richard Davison and Julia Steinberg for valuable discussions. We are also grateful to Mark Mezei for pointing out a correction to our chaos bound in an earlier version of the manuscript. C.Z. acknowledges financial support from the Harvard-MIT Center for Ultracold Atoms through NSF Grant No. PHY-1734011. The work of B.S. is supported in part by the AFOSR under grant number FA9550-19-1-0360.
\end{acknowledgments}

\bibliographystyle{apsrev4-2}
\bibliography{references}

\appendix

\section{Energy current}
\label{sec:appendixA}

The current flowing from $u$ to $v$ across the edge $(u, v)$ is given by 

\begin{equation}
    j_u = i\sum_{u'}A_{u'u}[H^{u'u}, H^{uv}],
\end{equation}
where $H^{uv} = H_0^u/d_u+H_0^v/d_v+H_1^{uv}$ is the on-bond Hamiltonian. The commutator simplifies to

\begin{equation}
\begin{split}
    j_u &= i\sum_{u'\neq v}A_{u'u}\Big(\frac{1}{d_u}\left([H_0^u, H_1^{uv}] -[H_0^u, H_1^{uu'}]\right) \\
    &+ [H_1^{u'u},H_1^{uv}]\Big).
    \label{eq:current}
\end{split}
\end{equation}
One can easily check using the anti-commutation relations that the commutator between a product of $m$ Majorana fermions and a product of $n$ Majorana fermions is proportional to $1-(-1)^{mn-l}$, where $l$ is the number of identical fermions in common between the two terms.  Hence the commutator vanishes unless $l$ is odd. Furthermore, in the large $N$ limit, the leading order contribution comes from the commutator where the terms have a single fermion in common, with other contributions suppressed by factors of $1/N$. 

To compute the non-equilibrium expectation value of any operator $O$ we employ the generating functional~\cite{kamenev2011field}
\begin{equation}
    \langle O(t) \rangle = \lim\limits_{\eta\to0} \frac{i}{2} \frac{\delta Z[\eta]}{\delta \eta(t)} = \lim\limits_{\eta\to0} \frac{i}{2} \frac{\delta \ln(Z[\eta])}{\delta \eta(t)} = \lim\limits_{\eta\to0} \frac{i}{2} \frac{\delta iS_\eta}{\delta \eta(t)},
\end{equation}
where $Z[\eta]$ is the generating functional with an additional source term $\eta(t)O(t)$. We also used the fact that $Z[\eta\to0] = 1$ in the Keldysh formalism~\cite{kamenev2011field}, and denoted by $S_\eta$ the part of the action that depends on the source term. The factor of $1/2$ accounts for the fact that $t$ can belong to either the positive $\mathcal{C}^+$ or negative $\mathcal{C}^-$ branch of the contour. Without loss of generality, we will assume that $t$ lives on $\mathcal{C}^+$ and ignore the factor of $1/2$.

Let us focus on the first commutator in \eqref{eq:current}. In order to find $S_\eta$, we have to perform the disorder averaging over couplings in the path integral, which results in a standard Gaussian integral~\cite{sarosi2017}. Its contribution to the action is given by

\begin{widetext}
\begin{equation}
    iS_\eta = -\frac{i^{q_S+q_I}}{2}NJ_0^2J_1^2 \int_{\mathcal{C}} \diff t_1 \diff t_2 G_u(t_1, t_2) G_u(t_1, t)^{q_S-1} G_u(t_2, t)^{\frac{q_I}{2}-1} G_v(t_2, t)^{\frac{q_I}{2}} \eta(t).
\end{equation}
Note that due to the causal structure, we only have to consider times $t_1, t_2 \leq t$. However, they can belong to either of the two contour branches, resulting in eight different orderings, which we group into four terms as follows:

\begin{align}
    j_{++}^{uv} &= \int_{-\infty}^{t} \diff t_2 \int_{-\infty}^{t_2} \diff t_1 G_u^<(t_1, t_2) \left(G_u^<(t_1, t) G_u^<(t_2, t)\right)^{\frac{q_I}{2}-1} \left(G_u^<(t_1, t)^{q_S-\frac{q_I}{2}} G_v^<(t_2, t)^{\frac{q_I}{2}} - G_v^<(t_1, t)^{\frac{q_I}{2}} G_u^<(t_2, t)^{q_S-\frac{q_I}{2}} \right),\\
    j_{+-}^{uv} &= \int_{-\infty}^{t} \diff t_2 \int_{-\infty}^{t_2} \diff t_1 G_u^<(t_1, t_2) \left(G_u^<(t_1, t) G_u^>(t_2, t)\right)^{\frac{q_I}{2}-1} \left(G_v^<(t_1, t)^{\frac{q_I}{2}} G_u^>(t_2, t)^{q_S-\frac{q_I}{2}}-G_u^<(t_1, t)^{q_S-\frac{q_I}{2}} G_v^>(t_2, t)^{\frac{q_I}{2}} \right),\\
    j_{-+}^{uv} &= \int_{-\infty}^{t} \diff t_2 \int_{-\infty}^{t_2} \diff t_1 G_u^>(t_1, t_2) \left(G_u^>(t_1, t) G_u^<(t_2, t)\right)^{\frac{q_I}{2}-1} \left(G_v^>(t_1, t)^{\frac{q_I}{2}} G_u^<(t_2, t)^{q_S-\frac{q_I}{2}} - G_u^>(t_1, t)^{q_S-\frac{q_I}{2}} G_v^<(t_2, t)^{\frac{q_I}{2}}\right),\\
    j_{--}^{uv} &= \int_{-\infty}^{t} \diff t_2 \int_{-\infty}^{t_2} \diff t_1 G_u^>(t_1, t_2) \left(G_u^>(t_1, t) G_u^>(t_2, t)\right)^{\frac{q_I}{2}-1} \left(G_u^>(t_1, t)^{q_S-\frac{q_I}{2}} G_v^>(t_2, t)^{\frac{q_I}{2}} - G_v^>(t_1, t)^{\frac{q_I}{2}} G_u^>(t_2, t)^{q_S-\frac{q_I}{2}} \right).
\end{align}
Notice that we used the identities in \eqref{eq:contour_identity} to write the answers exclusively in terms of greater or lesser Green's functions. Additionally, using \eqref{eq:gtr_lsr}, we obtain that $j_{++}^{uv} = (j_{--}^{uv})^*$ and $j_{+-}^{uv} = (j_{-+}^{uv})^*$. The expectation value of the commutator is given by the sum of the four terms above, with the appropriate proportionality constant
%the relationship between the lesser and greater Green's functions in

\begin{equation}
    \langle [H_0^{u}, H_1^{uv}] \rangle = -\frac{i^{q_S+q_I+1}}{2}J_0^2J_1^2(j_{++}^{uv}+j_{+-}^{uv}+j_{-+}^{uv}+j_{--}^{uv}) = -i^{q_S+q_I+1}J_0^2J_1^2\Re(j_{++}^{uv}+j_{+-}^{uv}).
\end{equation}

A similar expression can be obtained for $[H_1^{u'u}, H_1^{uv}]$ if we define
\begin{align}
    j_{++}^{u'uv} &= \int_{-\infty}^{t} \diff t_2 \int_{-\infty}^{t_2} \diff t_1 G_u^<(t_1, t_2) \left(G_u^<(t_1, t) G_u^<(t_2, t)\right)^{\frac{q_I}{2}-1} \left(G_{u'}^<(t_1, t)^{\frac{q_I}{2}} G_v^<(t_2, t)^{\frac{q_I}{2}} - G_v^<(t_1, t)^{\frac{q_I}{2}} G_{u'}^<(t_2, t)^{\frac{q_I}{2}} \right),\\
    j_{+-}^{u'uv} &= \int_{-\infty}^{t} \diff t_2 \int_{-\infty}^{t_2} \diff t_1 G_u^<(t_1, t_2) \left(G_u^<(t_1, t) G_u^>(t_2, t)\right)^{\frac{q_I}{2}-1} \left(G_v^<(t_1, t)^{\frac{q_I}{2}} G_{u'}^>(t_2, t)^{\frac{q_I}{2}}-G_{u'}^<(t_1, t)^{\frac{q_I}{2}} G_v^>(t_2, t)^{\frac{q_I}{2}} \right),\\
    j_{-+}^{u'uv} &= \int_{-\infty}^{t} \diff t_2 \int_{-\infty}^{t_2} \diff t_1 G_u^>(t_1, t_2) \left(G_u^>(t_1, t) G_u^<(t_2, t)\right)^{\frac{q_I}{2}-1} \left( G_v^>(t_1, t)^{\frac{q_I}{2}} G_{u'}^<(t_2, t)^{\frac{q_I}{2}}-G_{u'}^>(t_1, t)^{\frac{q_I}{2}} G_v^<(t_2, t)^{\frac{q_I}{2}} \right),\\
    j_{--}^{u'uv} &= \int_{-\infty}^{t} \diff t_2 \int_{-\infty}^{t_2} \diff t_1 G_u^>(t_1, t_2) \left(G_u^>(t_1, t) G_u^>(t_2, t)\right)^{\frac{q_I}{2}-1} \left(G_{u'}^>(t_1, t)^{\frac{q_I}{2}} G_v^>(t_2, t)^{\frac{q_I}{2}} - G_v^>(t_1, t)^{\frac{q_I}{2}} G_{u'}^>(t_2, t)^{\frac{q_I}{2}} \right),
\end{align}
where only the index of the terms in brackets changed from $u$ to $u'$ and $q_S$ was changed to $q_I$. The expectation value of the commutator is now given by

\begin{equation}
    \langle [H_1^{u'u}, H_1^{uv}] \rangle = -\frac{i}{2}J_1^4(j_{++}^{u'uv}+j_{+-}^{u'uv}+j_{-+}^{u'uv}+j_{--}^{u'uv}) = -iJ_1^4\Re(j_{++}^{u'uv}+j_{+-}^{u'uv})
\end{equation}
Putting everything together using \eqref{eq:current}, we arrive at the following expression for the current:

\begin{equation}
    \langle j_u \rangle =  J_1^2\sum_{u'\neq v}A_{u'u} \left(i^{q_S+q_I}\frac{J_0^2}{d_u}\Re(j_{++}^{uv}+j_{+-}^{uv}) - i^{q_S+q_I}\frac{J_0^2}{d_u}\Re(j_{++}^{uu'}+j_{+-}^{uu'}) + J_1^2\Re(j_{++}^{u'uv}+j_{+-}^{u'uv})\right).
    \label{eq:current_general}
\end{equation}
\end{widetext}

It is important to mention that in non-equilibrium situations where the Green's functions are only weakly perturb from their equilibrium values $G_u^> = G^{(0)}+F_u$ with $|F_u|\ll|G^{(0)}|$ and $q_S=q_I$, there are significant simplifications to the formula above

\begin{align}
    j_{+\pm}^{u'uv} &= j_{+\pm}^{uv}- j_{+\pm}^{uu'}, \\
    \langle j_u \rangle &= \frac{J_1^2(J_0^2+d_u J_1^2)}{d_u}\sum_{u'\neq v}A_{u'u} \Re(j_{++}^{u'uv}+j_{+-}^{u'uv}).
    \label{eq:current_final}
\end{align}
We verified numerically that this condition applies to all the non-equilibrium setups studied in this paper and therefore use \eqref{eq:current_final} throughout, except for the cases when $q_S \neq q_I$, where the more general \eqref{eq:current_general} applies.  

\section{Chaos bound on diffusion}
\label{sec:appendixB}

In this section, we review the many-body chaos properties of the SYK model. Chaos is fundamentally related to energy fluctuations and diffusion, since the local energy characterizes the rate of change of the quantum phase, and phase decoherence leads to chaos~\cite{davison2017}. We begin by introducing the out-of-time-order correlation function (OTOC), which has been widely used as a measure of chaos in quantum systems~\cite{larkin1969,shenker2014,shenker2015,maldacena2016chaos,maldacena2016,bryce2021}. Following~\cite{maldacena2016,gu2017diffusion}, we define the regularized OTOC in real time

\begin{equation}
    C(x, t_1, t_2) = \frac{1}{N^2}\sum_{i, j=1}^N \Tr[y\psi_j^x(t_1)y\psi_i^0(0)y\psi_j^x(t_2)y\psi_i^0(0)],
\end{equation}
where $y=e^{-\beta H/4}/Z^{1/4}$ evenly spaces the fermionic fields along the thermal circle. The OTOC measures how sensitive the system is to an initial local perturbation created by the operator $\psi_i^0(0)$, thus characterizing a quantum analog for the butterfly effect~\cite{shenker2014}. The leading order contribution to the OTOC comes from a time independent constant equal to the disconnected correlator $\mathcal{F}_d$~\cite{mezei2020}. The next order contribution comes from contracting $\psi_i$ with $\psi_j$ and is of order $1/N$ at early times

\begin{equation}
    C(x, t_1, t_2) = \mathcal{F}_d-\frac{\mathcal{F}(x, t_1, t_2)}{N}.
\end{equation}
For chaotic systems with a large
hierarchy between thermalization and scrambling, we expect $\mathcal{F}$ to grow exponentially as $e^{\lambda_Lt}$, where $t_1=t_2=t$ is in the Lyapunov regime $\beta\lesssim t\lesssim \beta\ln N$~\cite{mezei2020}. Here $\lambda_L$ is the Lyapunov exponent determining the scrambling rate~\cite{shenker2015}. For a single SYK cluster (with no spatial dependence), $\mathcal{F}(t_1, t_2)$ is determined by summing over a set of ladder diagrams~\cite{maldacena2016}, leading to the self-consistency equation 

\begin{equation}
    \mathcal{F}(t_1, t_2) = \int_{-\infty}^\infty \diff t_3 \diff t_4 K^R(t_1, t_2, t_3, t_4) \mathcal{F}(t_3, t_4),
\end{equation}
where $K_R$ is the retarded kernel 

\begin{equation}
\begin{split}
    K^R(t_1, t_2, t_3, t_4) &= \\
    &\hspace{-1.7cm} -(q-1)J^2G^R(t_1-t_3)G^R(t_2-t_4)G^W(t_3-t_4)^{q-2}. 
    \label{eq:kernel}
\end{split}
\end{equation}
The functions $G^R$ and $G^W$ are the retarded and Wightman Green's functions. The Wightman propagator is related to the spectral function in frequency space~\cite{guo2019}
%For fermionic systems,

\begin{equation}
    G^W(\omega) = \frac{A(\omega)}{2\cosh(\beta\omega/2)}.
    \label{eq:wightman}
\end{equation}

To determine the Lyapunov exponent, we follow the prescription introduced in Ref.~\cite{gu2019}. We define a variant of the kernel with a parameter $\alpha < 0$

\begin{equation}
    K_\alpha^R(t, t') = \int_{-\infty}^\infty \diff s e^{\alpha s} K^R\left(s+\frac{t}{2}, s-\frac{t}{2}, \frac{t'}{2}, -\frac{t'}{2}\right).
    \label{eq:kernel_alpha}
\end{equation}
We can view this operator as a matrix with its largest eigenvalue denoted by $k_R(\alpha)$. Then the Lyapunov exponent is determined by the equation $k_R(-\lambda_L)=1$. This is equivalent to solving for $\mathcal{F}$ as an eigenvector of the kernel $K^R$ with eigenvalue one. The $(0+1)$-d SYK model is known to saturate a bound on the Lyapunov exponent at low temperatures $\lambda_L = 2\pi/\beta$~\cite{maldacena2016,maldacena2016chaos}. 

The one-dimensional SYK chain allows us to study the chaos dynamics in space. To characterize the spatial propagation in a translation-invariant system, it is convenient to first introduce the Fourier transform $\mathcal{F}(p, \omega)$. As derived in Ref.~\cite{guo2019} using the ladder identity, the OTOC has a pole in both frequency and momentum space 

%\begin{equation}
%    \mathcal{F}(p, \omega) = \int_{-\infty}^\infty \diff x \diff t e^{-ipx+i\omega t} \mathcal{F}(x, t). 
%\end{equation}

\begin{equation}
    \mathcal{F}(p, \omega) \sim \frac{1}{\cos(\lambda_L(p)\beta/4)}\frac{1}{\omega-i\lambda_L(p)},
\end{equation}
where $\lambda_L(p)$ is the momentum-dependent Lyapunov exponent. By performing an inverse Fourier transform back to real space, we find~\cite{guo2019} 

\begin{equation}
    \mathcal{F}(x, t) \sim \int_{-\infty}^\infty \frac{\diff p}{2\pi} \frac{e^{\lambda_L(p)t+ipx}}{\cos(\lambda_L(p)\beta/4)}. 
\end{equation}
At large distances and times, this integral can be evaluated using a saddle point approximation. Depending on the parameters of our model, the integral can either pick up a contribution solely from the saddle point $p_s$, or from both the saddle point and the momentum-space pole $p_1$, both of which are located on the imaginary axis $p_{s, 1}=i|p_{s, 1}|$~\cite{gu2019,guo2019}. If $|p_s|<|p_1|$, the OTOC receives a contribution only from the saddle point 

\begin{equation}
    \mathcal{F}(x, t) \sim e^{\lambda_L(p_s)t+ip_s|x|} = e^{\lambda_L(p_s)(t-|x|/v_s)},
\end{equation}
where $v_s \equiv \lambda_L(p_s)/|p_s|$. Conversely, if $|p_s|>|p_1|$, the OTOC receives a dominant contribution from the pole, resulting in a wave-front that propagates with a maximal chaos rate $\lambda_L(p_1)=2\pi/\beta$

\begin{equation}
    \mathcal{F}(x, t) \sim e^{\lambda_L(p_1)t+ip_1|x|} = e^{\frac{2\pi}{\beta}(t-|x|/v_1)},
\end{equation}
where $v_1=2\pi/\beta |p_1|$. We can now define the butterfly velocity~\cite{gu2019,guo2019,choi2021}
%as the boundary of the region in which the OTOC grows 
\begin{equation}
    v_B = \begin{cases} v_s &\mbox{if } |p_s|<|p_1| \\
            v_1 & \mbox{if } |p_s| > |p_1| \end{cases}
\end{equation}
Physically, $v_B$ represents the growth rate of the region where operators have large anti-commutators with the initial $\psi_i^0(0)$. We can use the butterfly velocity to define two chaos diffusion constants, $\beta v_B^2/2\pi$ and $v_B^2/\lambda_L$, which are known to be closely related to the energy diffusion constant in strange metals~\cite{blake2016_1,blake2016_2,blake2017,guo2019,hartnoll2014,davison2017,gu2017,gu2017diffusion,choi2021,hartman2017}.

\begin{figure}[htp]
	\begin{center}
	\includegraphics[width = \columnwidth]{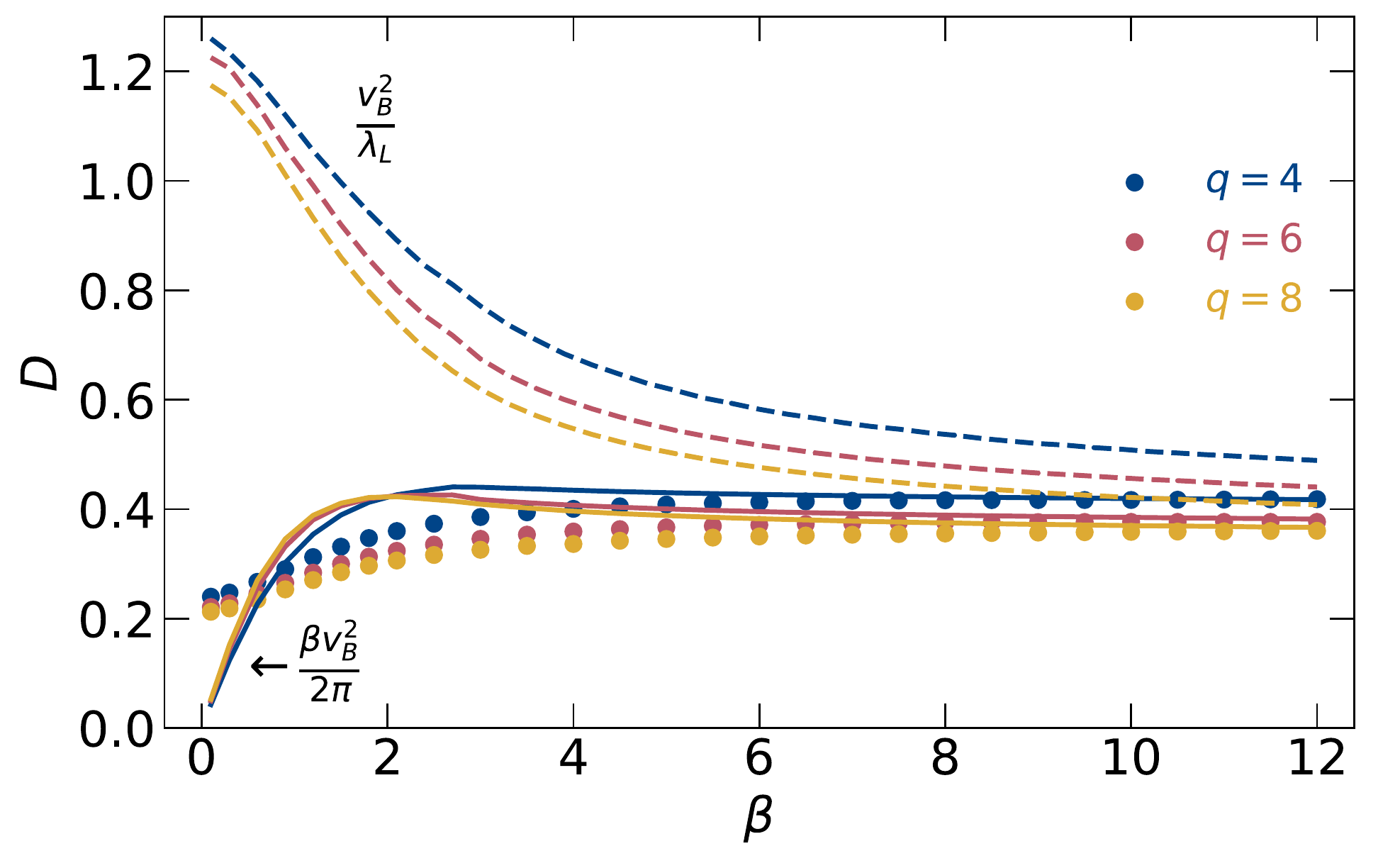}
	\caption{Temperature dependence of the energy diffusion constants (circles) and the chaos bounds $\beta v_B^2/2\pi$ (solid lines) and $v_B^2/\lambda_L$ (dashed lines) for different values of $q$. All three quantities agree in the conformal limit.}
	\label{fig:fig9}
	\end{center}
\end{figure}

In the remainder of this section, we will compute the value of $v_B$ and show that although $\beta v_B^2/2\pi$ closely follows the energy diffusion coefficient $D$ at low temperatures, the larger quantity $v_B^2/\lambda_L$ provides a true upper bound on energy transport. Note that we only have to find the momenta $p_{s, 1}$. We follow the procedure in Ref.~\cite{gu2019}. For the SYK chain, the retarded kernel factorizes in momentum space $K^R(p) = s(p)K^R$, where $s(p)=1+\frac{qJ_1^2}{2(q-1)J^2}(\cos(p)-1)$ is the spatial kernel~\cite{gu2017diffusion} and $K^R$ is the kernel for a single cluster with effective coupling $J$ defined in \eqref{eq:kernel}. Hence the eigenvalues of the kernel simply get rescaled $k_R(p, \alpha) = s(p)k_R(\alpha)$. The momentum-dependent Lyapunov exponent can be obtained by solving the equation $k_R(p, -\lambda_L(p))=1$. The location of the saddle can be found by solving $\lambda_L(p_s)=p_s\lambda_L^{'}(p_s)$, while $p_1$ is the momentum at which the Lyapunov exponent attains its maximum value $\lambda_L(p_1)=2\pi/\beta$. Note that for $q=2$, the model is non-chaotic and the Lyapunov exponent vanishes~\cite{garcia2018,haque2019}.

We numerically diagonalize the kernel and compute $\beta v_B^2/2\pi$ and $v_B^2/\lambda_L$ as a function of temperature. Our results are shown in \figref{fig:fig9}. At infinite temperature, both $v_B$ and $\lambda_L$ approach a constant. Therefore, their ratio $v_B^2/\lambda_L$ also approaches a constant, while $\beta v_B^2/2\pi$ decays to zero. At high temperatures, we have $D > \beta v_B^2/2\pi$, which was previously observed in a large $q$ expansion of this model~\cite{choi2021}. At low temperatures, on the other hand, we find $\lambda_L = 2\pi/\beta$ and all three quantities converge to the same diffusion constant $D=v_B^2/\lambda_L=\beta v_B^2/2\pi$, as was previously shown analytically~\cite{gu2017diffusion}. This indicates that the SYK chain is maximally chaotic in the conformal limit. This remarkable result is a consequence of the fact that the same reparameterization degrees of freedom are responsible both for energy diffusion and the OTOC chaos dynamics~\cite{maldacena2016,gu2017diffusion}.

Our results also show that $D\leq v_B^2/\lambda_L$ at all temperatures, suggesting that chaos upper bounds energy diffusion. Moreover, we see that the weaker bound $D\leq\beta v_B^2/2\pi$ is violated at high temperatures~\cite{choi2021}. In was previously conjectured that there is a fundamental lower bound on transport in incoherent metals $D\geq v^2\tau$, where $v$ and $\tau$ are the characteristic velocity and relaxation time respectively~\cite{hartnoll2014}. It was later suggested that many-body chaos provides a natural choice for these quantities: $v=v_B$ and $\tau=1/\lambda_L$~\cite{blake2016_1,blake2016_2,blake2017}. However, we see that the inequality is precisely reversed in our case, as was also found in other systems~\cite{gu2017,lucas2016,davison2017,choi2021}. This indicates that $v_B$ and $\lambda_L$ are not always the appropriate scales, and there is no simple bound relating transport and chaos in all incoherent metals~\cite{hartman2017}.

\section{Exact calculations of the diffusion constant}
\label{sec:appendixC}

In \secref{sec:ness_ansatz_dc} we introduced a simple ansatz for the non-equilibrium contribution to the Green's function $F(t)=\diff G^{(0)}(t)/\diff\beta$. Both the current and energy gradient have a relatively simple form in terms of $F(t)$. In what follows, we will consider the special cases where it is possible to compute the equilibrium Green's function $G^{(0)}(t)$, and hence $F(t)$, analytically. This will lead to closed-form expressions for the diffusion constant $D$ in various limits. In particular, we will consider the case of $q=2$, as well as the large $q$ approximation, in the limit of zero and infinite temperatures. For the zero-temperature limit, we will recover precisely the conformal answer in \eqref{eq:inf_beta_theory}. Our results are summarized in \tabref{tab:summary}.

\begin{table}[h]
  \centering
  \caption{\label{tab:summary} Diffusion constant computed exactly in various limits.}
  \begin{ruledtabular}
    \begin{tabular}{llll}
      Limit & $q=2$ & large $q$ & $q=\infty$\\
      \hline
      $\beta\to0$ & $\frac{32}{15\pi}\frac{J_1^2}{J}$ & $\frac{q}{q-2}\frac{\J_1^2}{3\J}$ & $\frac{\J_1^2}{3\J}$\\
      $\beta\to\infty$ & $\frac{\J_1^2}{\J}$ & $\frac{q}{q-2}\frac{\pi}{6}\frac{\J_1^2}{\J}$ & $\frac{\pi}{6}\frac{\J_1^2}{\J}$ \\
    \end{tabular}
  \end{ruledtabular}
\end{table}

\subsection{\texorpdfstring{$q=2$}{q=2} limit}

The SYK Hamiltonian for $q=2$ is equivalent to a random hopping model of free Majorana fermions, which can be solved exactly~\cite{maldacena2016,eberlein2017}. The spectral function is given by 

\begin{equation}
    A(\omega) = \frac{2}{J}\sqrt{1-\left(\frac{\omega}{2J}\right)^2} \quad \mathrm{for} \ |\omega|<2J.
\end{equation}
In equilibrium, the greater Green's function can be obtained from the fluctuation-dissipation theorem 

\begin{equation}
    G^{(0)}(\omega) = \frac{A(\omega)}{1+e^{-\beta\omega}},
\end{equation}
followed by an inverse Fourier transform 

\begin{equation}
\begin{split}
    G^{(0)}(t) &= \int_{-\infty}^\infty \frac{\diff\omega}{2\pi}e^{-i\omega t}G^{(0)}(\omega)\\
    &\hspace{-0.75cm} = -\frac{i}{2Jt}B(2Jt) - \frac{1}{\pi J}\int_{-2J}^{2J}\diff\omega \frac{\sin(\omega t)}{1+e^{-\beta\omega}}\sqrt{1-\left(\frac{\omega}{2J}\right)^2},
\end{split}
\end{equation}
where $B$ is the Bessel function of the first kind. Finally, we arrive at 

\begin{equation}
    F(t) = -\frac{2}{\pi J}\int_{0}^{2J}\diff\omega \frac{\omega\sin(\omega t)e^{-\beta\omega}}{(1+e^{-\beta\omega})^2}\sqrt{1-\left(\frac{\omega}{2J}\right)^2}.
\end{equation}
For $q=2$, the energy gradient in \eqref{eq:grad_E} takes the form 

\begin{equation}
    \nabla E = J\int_0^\infty \diff t \frac{F(t)}{t}B(2Jt),
\end{equation}
and the current in \eqref{eq:j} becomes

\begin{equation}
    j = -J_1^2\int_0^\infty \diff t \int_t^\infty \diff t' \frac{F(t')}{t(t'-t)} B(2Jt)B(2J(t'-t)).
\end{equation}

These integrals can be performed analytically in the $\beta\to\infty$ limit. To leading order in $1/\beta$, we find $\nabla E = -\pi/(6J\beta^3)$ and $j=\pi J_1^2/(6J^2\beta^3)$, concluding that
\begin{equation}
    D = \frac{J_1^2}{J} \quad (q=2, \beta\to\infty).
\end{equation}
This matches exactly the conformal answer in \eqref{eq:inf_beta_theory}, since $\alpha_K = \pi$ for the $q=2$ theory~\cite{maldacena2016}.

On the other hand, in the $\beta\to0$ limit, we find $\nabla E = -J^2/8$ and $j=4J_1^2J/15\pi$. This leads to
\begin{equation}
    D = \frac{32}{15\pi}\frac{J_1^2}{J} \quad (q=2, \beta\to 0),
\end{equation}
in agreement with the results presented in \figref{fig:fig4}(a). 

\begin{widetext}
\subsection{\texorpdfstring{$q\to\infty$}{Infinite q} limit}

We now return to the large $q$ analysis of \secref{sec:large_q_results} and take the limit of infinite $q$, while keeping $\beta$ arbitrary. We already have an expression for $\nabla E$ in \eqref{eq:grad_E_1}, so we only have to compute the current. When we take $q\to\infty$, while keeping $\J_0$ and $\J_1$ constant, \eqref{eq:j_general} simplifies to

\begin{equation}
    j = \frac{\J_1^2\J^2}{2q^2}\int_{0}^{\infty} \diff t \int_{t}^{\infty} \diff t' \left(\re\left[e^{g^{(0)}(t)}f(t)\right]\im\left[e^{g^{(0)}(t')}\right]- \re\left[e^{g^{(0)}(t)}\right]\im\left[e^{g^{(0)}(t')}f(t')\right] \right).
    \label{eq:j_inf_q}
\end{equation}
We will use the ODE for $f(t)$ (\eqref{eq:ODE}) and the explicit form of $g^{(0)}(t)$ (\eqref{eq:g_0}) repeatedly to simplify the equation above. First, using the differential equation, we find

\begin{align}
    \int_{t}^{\infty} \diff t' \im\left[e^{g^{(0)}(t')}\right] &= -\frac{1}{2\J^2}\int_{t}^{\infty} \diff t' \im\left[g^{(0)}(t')''\right] = \frac{\im\left[g^{(0)}(t)'\right]}{2\J^2},\\
    \int_{t}^{\infty} \diff t' \im\left[e^{g^{(0)}(t')}f(t')\right] &= -\frac{1}{2\J^2}\int_{t}^{\infty} \diff t' \im\left[f''(t')\right] = \frac{\im\left[f'(t)\right]}{2\J^2}.
\end{align}
Next, we perform an integration by parts using the fact that $\re\left[g^{(0)}(0)'\right] = 0$ and $\im\left[f'(0)\right] =0$

\begin{equation}
    \int_{0}^{\infty} \diff t \re\left[e^{g^{(0)}(t)}\right]\im\left[f'(t)\right] = -\frac{1}{2\J^2}\int_{0}^{\infty} \diff t \re\left[g^{(0)}(t)''\right]\im\left[f'(t)\right] = - \int_{0}^{\infty} \diff t \re\left[g^{(0)}(t)'\right]\im\left[e^{g^{(0)}(t)}f(t)\right].
\end{equation}
Plugging this back into \eqref{eq:j_inf_q} we get

\begin{equation}
\begin{split}
    j &= \frac{\J_1^2}{4q^2}\int_{0}^{\infty} \diff t \left(\re\left[e^{g^{(0)}(t)}f(t)\right]\im\left[g^{(0)}(t)'\right]+ \re\left[g^{(0)}(t)'\right]\im\left[e^{g^{(0)}(t)}f(t)\right]\right) \\
    &= \frac{\J_1^2}{4q^2}\int_{0}^{\infty} \diff t \im\left[e^{g^{(0)}(t)}g^{(0)}(t)'f(t)\right]  = -\frac{\J_1^2}{4q^2}\int_{0}^{\infty} \diff t \im\left[e^{g^{(0)}(t)}f'(t)\right],
\end{split}
\end{equation}
\end{widetext}
where in the last step we used integration by parts and $e^{g^{(0)}(\infty)} = f(0) = 0$. The last integral can be evaluated by substituting the explicit formula for $f(t)$ from \eqref{eq:f_large_q}

\begin{equation}
    j = \frac{\J_1^2}{12q^2}\left(\frac{\pi v}{2}\sin\left(\pi v\right)+2\cos^2\left(\frac{\pi v}{2}\right)\right).
\end{equation}
Finally, combining this with \eqref{eq:grad_E_1} yields 

\begin{equation}
    D = \frac{\J_1^2}{3\J}\left(\frac{\pi v}{2}\sin(\frac{\pi v}{2})+\cos(\frac{\pi v}{2})\right) \quad (q\to\infty).
    \label{eq:infinite_q}
\end{equation}
Furthermore, we can investigate the different temperature limits. If $\beta\to0$, then $v\to0$ and 

\begin{equation}
    D = \frac{\J_1^2}{3\J} \quad (q\to\infty, \beta\to0).
\end{equation}
On the other hand, if $\beta\to\infty$, then $v\to1$ and 

\begin{equation}
    D = \frac{\pi}{6}\frac{\J_1^2}{\J} \quad (q\to\infty, \beta\to\infty).
\end{equation}
This again agrees with the conformal limit in \eqref{eq:inf_beta_theory}, since $\alpha_K=3$ at infinite $q$~\cite{maldacena2016}. 

\subsection{Finite \texorpdfstring{$q$}{q} corrections}

In our derivation above, we took the infinite $q$ limit first, followed by a temperature limit. However, we can also reverse the order to find finite $q$ corrections to the zero- and infinite-temperature diffusion constants. For instance, consider taking the limit $\beta\to0$ first. In this case, $v\to0$ as well, but their ratio approaches a constant $v/\beta \to \J/\pi$. To leading order, we approximate

\begin{align}
    e^{g^{(0)}(t)} &= \frac{1}{\cosh^2(\J t)},\\
    f(t) &= -i\tanh(\J t).
\end{align}
We now have an elementary solution for $f(t)$, which allows us to directly evaluate the integral in \eqref{eq:j_general} 
%for arbitrary $q$

\begin{equation}
    j = \frac{\J_1^2}{2q(q-2)}\left(1-\frac{\sqrt{\pi}}{2}\frac{\Gamma(2-2/q)}{\Gamma(5/2-2/q)}\right),
\end{equation}
where $\Gamma$ denotes the gamma function. Combining this with $\nabla E = \J/2q^2$, we arrive at

%\frac{q}{q-2}\left(1-\frac{\sqrt{\pi}}{2}\frac{\Gamma(2-2/q)}{\Gamma(5/2-2/q)}\right) \frac{\J_1^2}{\J}\approx
\begin{equation}
    D =  \frac{q}{q-2}\frac{\J_1^2}{3\J} \quad (\beta\to0),
\end{equation}
to leading order in $1/q$. A similar calculation for the $\beta\to\infty$ limit results in the same prefactor 

\begin{equation}
    D =  \frac{q}{q-2}\frac{\pi}{6}\frac{\J_1^2}{\J} \quad (\beta\to\infty).
\end{equation}
These results match the numerical values in \figref{fig:fig4}(b).

\section{Solution of the KB equations with non-equilibrium AC ansatz}
\label{sec:appendixD}

In this section, we present the KB equations for the non-equilibrium contributions to the Green's functions in the case of AC transport and discuss how to solve them numerically. We start by substituting the ansatz in \eqref{eq:ac_ness_ansatz} into \eqref{eq:KB}, and expanding to first order in $F_{1,2}(t)$. All the functions can be written in terms of the new time variables $t$ and $T$. Moreover, after a change of variables, all the dependence on $T$ factors out into $\cos(\omega T)$ and $\sin(\omega T)$ terms. Since the KB equations hold for arbitrary $T$, the prefactors in front of $\cos(\omega T)$ and $\sin(\omega T)$ must vanish independently. Therefore, each KB equation leads to a set of two new equations as follows:

\begin{align}
    \partial_t F_1(t)+\frac{\omega}{2}F_2(t) &= 2i^qJ^2 I_1(F_1, F_2), \label{eq:KB_AC_1}\\
    \partial_t F_2(t)-\frac{\omega}{2}F_1(t) &= 2i^qJ^2 I_1(F_2, -F_1), \label{eq:KB_AC_2}\\
    -\partial_t F_1(t)+\frac{\omega}{2}F_2(t) &= 2i^qJ^2 I_2(F_1, F_2), \label{eq:KB_AC_3} \\
    -\partial_t F_2(t)-\frac{\omega}{2}F_1(t) &= 2i^qJ^2 I_2(F_2, -F_1), \label{eq:KB_AC_4}
\end{align}
where we define the integrals

\begin{widetext}
\begin{align}
    I_1(F_1, F_2) &= -\int_0^t \diff t' \left(a\cos(\omega t'/2)+b\sin(\omega t'/2)\right)G^{(0)}(t')\im\left[G^{(0)}(t-t')^{q-2}F_1(t-t')\right] \nonumber\\
    &+ \int_0^\infty \diff t' \left(a\cos(\omega t'/2)-b\sin(\omega t'/2)\right)\im\left[G^{(0)}(t')G^{(0)}(t+t')^{q-2}F_1(t+t')\right] \nonumber\\
    &- \int_0^t \diff t' \cos(\omega t'/2)F_1(t-t')\im\left[G^{(0)}(t')^{q-1}\right]+ \int_t^\infty \diff t' \cos(\omega t'/2)\im\left[G^{(0)}(t')^{q-1}F_1(t'-t)\right] \nonumber\\
    &-\int_0^t \diff t' \left(a\sin(\omega t'/2)-b\cos(\omega t'/2)\right)G^{(0)}(t')\im\left[G^{(0)}(t-t')^{q-2}F_2(t-t')\right] \nonumber\\
    &- \int_0^\infty \diff t' \left(a\sin(\omega t'/2)+b\cos(\omega t'/2)\right)\im\left[G^{(0)}(t')G^{(0)}(t+t')^{q-2}F_2(t+t')\right] \nonumber\\
    &+ \int_0^t \diff t' \sin(\omega t'/2)F_2(t-t')\im\left[G^{(0)}(t')^{q-1}\right]- \int_t^\infty \diff t' \sin(\omega t'/2)\im\left[G^{(0)}(t')^{q-1}F_2(t'-t)\right],
\end{align}

\begin{align}
    I_2(F_1, F_2) &= \int_0^t \diff t' \left(a\cos(\omega t'/2)-b\sin(\omega t'/2)\right)G^{(0)}(t-t')^{q-2}F_1(t-t')\im\left[G^{(0)}(t')\right] \nonumber \\
    &- \int_t^\infty \diff t' \left(a\cos(\omega t'/2)-b\sin(\omega t'/2)\right)\im\left[G^{(0)}(t')G^{(0)}(t'-t)^{q-2}F_1(t'-t)\right] \nonumber\\
    &+ \int_0^t \diff t' \cos(\omega t'/2)G^{(0)}(t')^{q-1}\im\left[F_1(t-t')\right]- \int_0^\infty \diff t' \cos(\omega t'/2)\im\left[G^{(0)}(t')^{q-1}F_1(t+t')\right] \nonumber\\
    &-\int_0^t \diff t' \left(a\sin(\omega t'/2)+b\cos(\omega t'/2)\right)G^{(0)}(t-t')^{q-2}F_2(t-t')\im\left[G^{(0)}(t')\right] \nonumber \\
    &+ \int_t^\infty \diff t' \left(a\sin(\omega t'/2)+b\cos(\omega t'/2)\right)\im\left[G^{(0)}(t')G^{(0)}(t'-t)^{q-2}F_2(t'-t)\right] \nonumber\\
    &+ \int_0^t \diff t' \sin(\omega t'/2)G^{(0)}(t')^{q-1}\im\left[F_2(t-t')\right]+ \int_0^\infty \diff t' \sin(\omega t'/2)\im\left[G^{(0)}(t')^{q-1}F_2(t+t')\right].
\end{align}
\end{widetext}
Note that the second equation in each set can be obtained from the first by changing $F_1\to F_2$ and $F_2\to -F_1$. To shorten the notation, we also introduced the constants
\begin{align}
    a &= (q-1)+q\frac{J_1^2}{J^2}\left(\frac{1}{2}\left(c+\frac{1}{c}\right)\cos\delta - 1\right), \\
    b &= \frac{q}{2}\frac{J_1^2}{J^2}\left(c-\frac{1}{c}\right)\sin\delta.
\end{align}

%Formally, this system of equations can be written as a homogeneous Fredholm equation of the second kind, whose solution can be found by casting it to an eigenvalue problem~\cite{press2007numerical}. However, since finding the kernel for this eigenvalue equation is non-trivial, we instead opt to iteratively solve the original system.

\begin{figure}[htp]
	\begin{center}
	\includegraphics[width = \columnwidth]{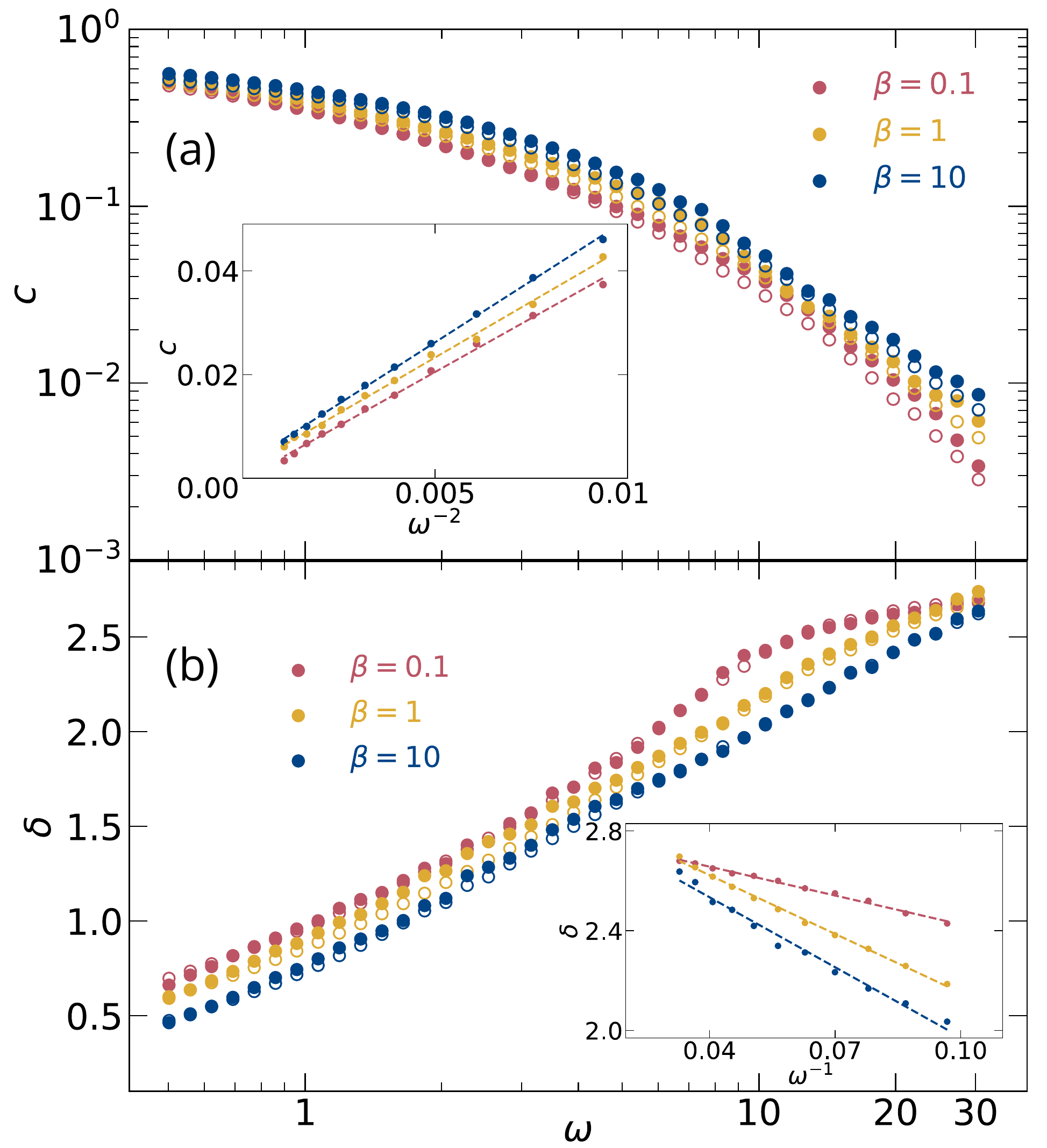}
	\caption{Frequency dependence of the (a) amplitude decay and (b) phase shift for the AC NESS solution. Filled circles represent numerical values extracted directly from the boundary-driven NESS. Empty circles are obtained by iteratively solving the system of equations in \appref{sec:appendixD}. The insets show high-frequency power-law fits to $c\sim \omega^{-2}$ and $\delta\sim \omega^{-1}$ respectively.}
	\label{fig:fig10}
	\end{center}
\end{figure}

We solve this system of equations iteratively. Recall that the parameters $(c, \delta)$ are not known a priori and have to be calculated during each iteration. In order to achieve convergence, we group the equations in a specific way. Our algorithm can be summarized as follows:

\begin{itemize}
    \item Solve the linear system of equations for $a$ and $b$ (or equivalently $c$ and $\delta$)
    \begin{align}
        \omega F_1(t) &= -2i^qJ^2\left(I_1(F_2, -F_1) + I_2(F_2, -F_1)\right), \\
        \omega F_2(t) &= 2i^qJ^2\left((I_1(F_1, F_2) + I_2(F_1, F_2)\right). 
    \end{align}
    \item Solve the integral-differential equations for $F_{1,2}(t)$
    \begin{align}
        \partial_t F_1(t) &= i^qJ^2\left(I_1(F_1, F_2) - I_2(F_1, F_2)\right), \\
        \partial_t F_2(t) &= i^qJ^2\left(I_1(F_2, -F_1) - I_2(F_2, -F_1)\right). 
    \end{align}
\end{itemize}
We start with an initial guess for $F_{1,2}(t)$ and $(c, \delta)$, and repeat the procedure above with a weighted update at each iteration until convergence (usually within $300$ iterations). It is worth mentioning that we can exactly recover the DC solution $F(t)=F_1(t)=F_2(t)$ by setting $\omega=\delta=0$ and $c=1$ in the previous equations. 

The iterative procedure is a lot faster to compute than the full time evolution of a boundary-driven chain and the results are more accurate. A sample solution for $\omega=5$ is shown in \figref{fig:fig7}(d). The values of $c$ and $\delta$ at different frequencies are plotted in \figref{fig:fig10}. We see a good agreement with the numerical results extracted from the boundary-driven NESS. Surprisingly, we find these parameters to be independent of $q$. At high frequencies, we find a quadratic dependence for the amplitude decay $c\sim\omega^{-2}$. This can be easily deduced from the scaling properties of \eqref{eq:KB_AC_1}; the left-hand side scales as $\omega$, while the right-hand side scales as $1/\omega c$ after integration. For the two sides to match, we must have $c\sim\omega^{-2}$. On the other hand, the phase scales as $\delta\sim \omega^{-1}$. At infinite frequency, we expect the consecutive sites to oscillate exactly out of phase with $\delta = \pi$.

\end{document}